%% file: main.tex
\documentclass[10pt,conference]{IEEEtran}
\pdfoutput=1
\usepackage{diagbox}
\usepackage{graphicx}
\usepackage{amsmath}
\usepackage{algpseudocode}
\usepackage{algorithm}
\usepackage{subfigure}

\newtheorem{theorem}{Theorem}
\newtheorem{proposition}{Proposition}
\newtheorem{lemma}{Lemma}
\newtheorem{corollary}{Corollary}
\newtheorem{example}{Example}

\begin{document}

\title{Demand-based Scheduling of Mixed-Criticality Sporadic Tasks on One Processor}
\author{\IEEEauthorblockN{Arvind Easwaran}
\IEEEauthorblockA{Nanyang Technological University, Singapore \\
Email: arvinde@ntu.edu.sg}}

\maketitle

\begin{abstract}
Strategies that artificially tighten high-criticality task deadlines in low-criticality behaviors have been successfully employed for scheduling mixed-criticality systems. Although efficient scheduling algorithms have been developed for implicit deadline task systems, the same is not true for more general sporadic tasks. In this paper we develop a new demand-based schedulability test for such general mixed-criticality task systems, in which we collectively bound the low- and high-criticality demand of tasks. We show that the new test strictly dominates the only other known demand-based test for such systems. We also propose a new deadline tightening strategy based on this test, and show through simulations that the strategy significantly outperforms all known scheduling algorithms for a variety of sporadic task systems.
\end{abstract}

\input{introduction}

\input{model}

\input{schedulability}

\input{simulations}

\section{Conclusions}
\label{sec:conclusions}

In this paper we derived a new demand-based schedulability test for general mixed-criticality sporadic task systems, and showed that it strictly dominates the existing demand-based test. A key insight used by this test is that by collectively considering low- as well as high-criticality demand in a time interval, the resulting bound can be far less pessimistic than independent bounds for low- and high-criticality demands. We also proposed a novel deadline tightening strategy based on this new test, and showed through simulations that it outperforms all known algorithms even under high load.

The test presented in this paper can only be used for constrained deadline task systems, and in particular, cannot be used when task deadlines are greater than minimum separation. In the future we will generalize our schedulability test for such arbitrary deadline sporadic task systems. 

%The \emph{speed-up} bound of an algorithm denotes the minimum speed necessary for the processor to guarantee that any task system feasible on a speed-1 processor is also schedulable by the algorithm. Since the problem of scheduling mixed-criticality workloads is known to be NP-Hard~\cite{BBD12}, the speed-up bound is a good metric for evaluating the performance of algorithms at least from a theoretical point of view. EDF-VD, proposed by Baruah \emph{et. al.}~\cite{BBA12}, is known to be speed-up optimal for task systems with two criticality levels and deadline equal to minimum separation (implicit deadline). Unfortunately, this algorithm cannot be used for scheduling more general sporadic task systems such as the one considered in this paper. In the future we plan to develop similar speed-up optimal algorithms for general sporadic task systems using the demand-based schedulability test presented here. To prove speed-up optimality of EDF-VD, a schedulability test that collectively considers demand in both low- and high-criticality behaviors was used, and therefore we believe that the demand-based test presented here can potentially lead to a speed-up optimal algorithm for general task systems.

The proposed test can be generalized to task systems with more than two criticality levels, by considering all possible time instants when the system switches between those multiple criticalities. But the complexity of this resulting test would be exponential in the number of criticality levels. In the future we also plan to develop a more computationally efficient demand-based test for such task systems.

\section*{Acknowledgment}

This work was supported by Start-up Grant, NTU, Singapore. The author would also like to thank his PhD student Xiaozhe Gu for helping with simulation experiments. 

\bibliographystyle{IEEEtran}
\bibliography{all}

\end{document}

%% file: introduction.tex
\section{Introduction}
\label{sec:introduction}

Scheduling problems for mixed-criticality real-time systems have received increasing attention in the recent past. These systems, first introduced by Vestal~\cite{Ves07}, comprise of real-time tasks with different criticality levels, all sharing the same hardware. One of the key requirements of such a system is the following: while all tasks should receive sufficient resources under normal operating conditions, tasks with higher criticality should be given preference over those with lower criticality when operating conditions deviate from the norm. Several formulations of what constitutes a deviation from the norm have been presented, including task execution beyond expected worst-case execution time (WCET)~\cite{Ves07}, scaling down of processor frequency~\cite{FGB13}, and task dispatch occurring more frequently than anticipated~\cite{BaBu11}. In this paper we focus on the formulation based on deviation from expected WCET.

Although several different mechanisms are available for bounding the WCET of a real-time task, they all introduce some level of pessimism to the bounds~\cite{WCET_survey09}. Additionally, some approaches such as those based on tests, may only provide a reasonable estimate of the WCET that is valid for almost all execution scenarios (occasional violations can occur). One may then wonder what level of confidence is desired when estimating the WCET of a real-time task. As a system designer whose focus is on efficient resource utilization, a smaller WCET estimate is desirable. On the other hand, as an authority responsible for ensuring the safety of the system (e.g., Federal Aviation Authority), an highly reliable estimate is desirable at least for the high criticality tasks. These seemingly contradictory WCET requirements have been formally captured using a list of WCET estimates for each real-time task, one estimate for each criticality level~\cite{Ves07}. While the actual execution time of each task remains below its lowest (least critical) WCET estimate, the system is assumed to be operating normally and all tasks are required to meet their deadlines. Any deviation from this norm has an implication that only certain high criticality tasks are required to meet their deadlines thereafter. It is important to note that the designer does not expect this deviation to occur during the system's lifetime (highly unlikely). Hence the deadline requirement after the deviation is only to convince the certification authorities that even in this unexpected situation, the critical tasks will continue to receive sufficient resources. We will use this mixed-criticality task model, and further details are presented in Section~\ref{sec:model}.

The sporadic task model, first introduced for non mixed-criticality systems~\cite{LiLa73}, is a generic model for capturing the real-time requirements of many event-driven systems including those with mixed-criticality such as avionics and automotive. It is therefore the model of choice in this work. 

When designing a scheduling algorithm for any real-time system, we believe the following properties are important. 
\begin{enumerate}
\item[\textbf{P1}:] Does the algorithm have a low run-time complexity for making scheduling decisions?
\item[\textbf{P2}:] Does the algorithm have an exact schedulability test (or if exact test is not feasible then tight test) with low time complexity?
\item[\textbf{P3}:] Does the algorithm, based on its schedulability test, successfully schedule a significantly large proportion of feasible\footnote{A task system is said to be feasible, if there exists some algorithm (clairvoyant or otherwise) that can successfully schedule it.} task systems?
\end{enumerate}      

\begin{figure*}
\centering
\begin{tabular}[t]{|c||c|c|c|c|}
\hline
\diagbox{Algorithm}{Property} & \shortstack{\textbf{P1} \\ Run-time Complexity} & \shortstack{\textbf{P2} \\ Test Complexity} & \shortstack{\textbf{P3} \\ Simulation Performance} \\ 
\hline
\hline
AMC~\cite{BBD11} & logarithmic~\cite{But05} & pseudo-polynomial~\cite{BBD11} &  Worst among the four \\
\hline
PLRS~\cite{GES11} & quadratic~\cite{GES11} & pseudo-polynomial~\cite{GES11} & Good at intermediate load, \\
& & & but poor at high load \\
\hline
GREEDY~\cite{EkYi12} & logarithmic~\cite{EkYi12} & pseudo-polynomial~\cite{EkYi12} & Better than AMC and PLRS at high load \\
\hline
ECDF (this paper) & logarithmic & pseudo-polynomial & Best (outperforms the rest and performance \\
& & & gap widens with increasing load)\\
\hline
\end{tabular}
\caption{Comparison of algorithms for general sporadic tasks. Conclusions regarding property \textbf{P3} are based on observations from simulations (see Section~\ref{sec:simulations}).}
\label{fig:algorithms_summary}
\end{figure*}

Property \textbf{P1} ensures that the scheduling algorithm has a low implementation overhead, so that platform resources can be utilized more efficiently for getting actual work done. Property \textbf{P2} enables off-line validation of the real-time requirements of the system in an efficient manner. For instance, certification authorities could use this schedulability test as one of the metrics in the certification process. Exactness (or tightness) of the test ensures that a large proportion of task systems schedulable by the algorithm are successfully identified by the test. Finally, a scheduling algorithm that satisfies property \textbf{P3} can more likely than not successfully schedule a system in practice. This property could be deduced either analytically (e.g., for optimal algorithms), or through carefully designed simulation experiments. Note that for general non mixed-criticality sporadic task systems it has been shown that determining feasibility is NP-Hard~\cite{BMR90}. This indicates that polynomial time schedulability tests may, in general, not lead to a good performance for property \textbf{P3}, and therefore we focus on pseudo-polynomial tests in this paper.

Some algorithms have been proposed for the scheduling of mixed-criticality sporadic task systems on a single processor (e.g.,~\cite{Ves07,BaVe08,LiBa10a,BBD11,GES11,BBG11,EkYi12}). In Figure~\ref{fig:algorithms_summary}, we summarize the performance of these algorithms in relation to the three properties described above. Vestal~\cite{Ves07} first proposed an algorithm based on Audsley's priority assignment strategy~\cite{Aud91}, which has since been dominated by another algorithm proposed by Baruah \emph{et. al.}~\cite{BBD11}. Assuming run-time support from the platform, this new algorithm adapts task priorities once a deviation from the norm is detected. This algorithm is denoted as \emph{AMC} for "Adaptive Mixed Criticality"  in this paper. As can be seen from Figure~\ref{fig:algorithms_summary}, AMC has a low run-time complexity\footnote{This complexity can be reduced depending on the kind of support available in the kernel as well as the number of different priority levels~\cite{But05}.}, and the schedulability test is pseudo-polynomial. But its performance in simulations is relatively poor, especially when system load is high (see Figures~\ref{fig:deadline_normal} and~\ref{fig:deadline_skewed} in Section~\ref{sec:simulations}).

Algorithms that assign different priorities to different jobs of the same task have also been proposed in the past~\cite{LiBa10a,GES11,EkYi12}. Li and Baruah~\cite{LiBa10a} proposed the "Own Criticality Based Priority" algorithm for sporadic task systems, and its run-time complexity was improved from pseudo-polynomial to quadratic by Guan \emph{et. al.}~\cite{GES11} when they proposed the "Priority List Reuse Scheduling" (PLRS) algorithm. Both pseudo-polynomial tests based on response-time analysis and polynomial tests based on system load have been proposed for OCBP and PLRS. But, similar to AMC, these algorithms also have relatively poor performance in simulations especially when the system load is high (see figures in Section~\ref{sec:simulations}). To overcome this performance gap in AMC, OCBP, and PLRS algorithms, Ekberg and Yi~\cite{EkYi12} proposed a greedy search algorithm (called GREEDY in this paper). This algorithm artificially tightens the deadline of high-criticality tasks when the system is still operating normally, so that when the system deviates from the norm, additional time is available to schedule the workload of these high-criticality tasks. Although GREEDY has a logarithmic run-time complexity and a schedulability test with pseudo-polynomial complexity, its performance in simulations is better than AMC and PLRS only at high system load. At intermediate load it is actually outperformed by PLRS (see Figure~\ref{fig:deadline_normal} in Section~\ref{sec:simulations}). Further, as can be seen from those figures, GREEDY is still not able to schedule a large fraction of potentially feasible sporadic task systems.

%Baruah \emph{et. al.}~\cite{BBA12} proposed an algorithm called "Earliest Deadline First - Virtual Deadlines (EDF-VD)" that also employs a deadline tightening strategy. Although this algorithm is known to be theoretically efficient, it is only applicable for the restricted class of implicit deadline tasks and cannot be used for the more general task systems considered in this paper. 
\textbf{Contributions.} In this paper we propose a new demand-based schedulability test for general mixed-criticality sporadic task systems, and prove that it strictly dominates the only other known demand-based test (the one used by GREEDY). The key contributing factor for this dominance is an approach to collectively bound the low- as well as high-criticality demand of tasks in any time interval, as opposed to independent low- and high-criticality bounds that were used in the previous test.

We also propose a deadline tightening strategy for high-criticality tasks that uses the improved test, and show through simulations that the resulting scheduler significantly outperforms all known schedulers. In fact, simulation results show that this performance gap widens with increasing system load. Further, simulations also show that the performance of the proposed strategy is comparable to exhaustive deadline search based on the test, when number of tasks in a system are small.
%indicating that the proposed strategy can also efficiently schedule many hard-to-schedule task systems. 
%In this deadline tightening strategy we look at time intervals for which the new test fails, and then identify a high-criticality task that has a \emph{carry-over} job with earliest deadline. A carry-over job is a job that is released and ready for execution when the system is operating normally, and remains unfinished even after the system has deviated from the norm. Hence we call this strategy Earliest Carry-over Deadline First (ECDF). We artificially tighten the deadline of this task so that the carry-over job is no longer able to execute when the system deviates from the norm, and therefore contributes reduced demand. 

%% file: model.tex
\section{System Model}
\label{sec:model}

A mixed-criticality sporadic task can be specified as $\tau_i = (T_i, D_i, L_i, \mathcal{C}_i)$, where $T_i$ denotes minimum separation between job releases, $D_i$ denotes deadline, $L_i$ denotes criticality level, and $\mathcal{C}_i$ is a list of WCET values. In this paper we assume that tasks have only two criticality levels, $LC$ denoting low-criticality and $HC$ denoting high-criticality. We also assume that tasks are constrained deadline, meaning $D_i \leq T_i$. Note that even for this restricted model, the mixed-criticality scheduling problem is known to be NP-Hard~\cite{BBD12}.

For a dual-criticality task $\tau_i$, $L_i \in \{ LC, HC \}$ and $\mathcal{C}_i = \{ C_i^L, C_i^H \}$, where $C_i^L$ denotes $LC$ WCET and $C_i^H$ denotes $HC$ WCET. We assume that $C_i^L \leq C_i^H$ for all tasks $\tau_i$. Task $\tau_i$ releases an infinite sequence of jobs, each job has a deadline that is $D_i$ time units after its release, and successive jobs are released with a minimum separation of $T_i$ time units. A sporadic task system is then represented by a set of such tasks and denoted as $\tau = \{ \tau_1, \ldots , \tau_n \}$. The real-time requirements of such a task system can be summarized as follows:
\begin{enumerate}
\item As long as no job executes for more than its $LC$ WCET ($C_i^L$), the system is regarded as exhibiting $LC$ behavior, and all job deadlines must be met.
\item If at some time instant a $HC$ job executes beyond its $LC$ WCET, the system is then regarded as exhibiting $HC$ behavior, and only $HC$ job deadlines are required to be met after this time instant.
\item If a $LC$ (likewise $HC$) job executes beyond its $LC$ (likewise $HC$) WCET, then the system is regarded as exhibiting erroneous behavior. Therefore, in our system model, it can be assumed without loss of generality that $C_i^L = C_i^H$ for a $LC$ task $\tau_i$.
\end{enumerate}

As long as all jobs are executing within their $LC$ WCETs, the system is behaving as expected by the designer, and all job deadlines are required to be met. When a $HC$ job executes for more than its $LC$ WCET, the system has deviated from the norm because a job executed for more time than expected. Although this scenario is highly unlikely, the authorities responsible for system safety want to ensure that even in this case at least the critical functionalities are operational. In other words, they want to ensure that at least all the $HC$ jobs will continue to meet their deadlines. This implies that all $LC$ job deadlines after the deviation can be safely ignored. If a scheduling algorithm can satisfy all the above requirements for task system $\tau$, then $\tau$ is said to be \emph{schedulable} by the algorithm. Note the problem of determining when to switch the system back to $LC$ behavior after a $HC$ behavior is beyond the scope of this paper, because it has no impact on the schedulability tests and algorithms that we derive. 
%One simple strategy would be to switch the system back when no $HC$ task is pending.

\textbf{Scheduling strategy}: When the system is in $LC$ behavior all the tasks will be scheduled using Earliest Deadline First (EDF) strategy, and when the system switches to $HC$ behavior, all the $LC$ jobs will be ignored thereafter and only the $HC$ tasks will be scheduled using EDF strategy. This strategy was first introduced by Baruah \emph{et. al.}~\cite{BBG11} in EDF-VD and is also used by the GREEDY algorithm~\cite{EkYi12}. To accommodate the sudden increase in demand of $HC$ tasks when the system switches behavior, both EDF-VD and GREEDY propose artificial tightening of $HC$ task deadlines in $LC$ system behavior. This ensures that when the behavior switch occurs, all the active $HC$ jobs have some amount of time left until their real deadline to execute any additional demand in the $HC$ behavior. In this scheduling strategy, the key aspect is the mechanism for determining artificial deadlines. Unlike EDF-VD, which determines deadlines by evenly distributing remaining $LC$ utilization among the $HC$ tasks, GREEDY performs a heuristic search in the solution space. In this paper, we propose a new strategy for determining these deadlines, details of which are presented in Section~\ref{sec:strategy}.

Let $D_i^L$ denote the deadline of task $\tau_i$ in $LC$ behavior, also denoted as \emph{tightened deadline} to distinguish it from the \emph{real deadline} $D_i$. Since deadlines of $LC$ tasks are unmodified, we have $D_i^L = D_i$ for such tasks. For $HC$ tasks, $D_i^L \leq D_i$ by definition. We also denote the set of all $LC$ tasks in $\tau$ by $\mathcal{L}_{\tau}$, and the set of all $HC$ tasks in $\tau$ by $\mathcal{H}_{\tau}$. 

The \emph{demand bound function} (dbf) of a task for a given time interval length is the maximum demand that the task can impose in any time interval of that length. For a mixed-criticality task $\tau_i$ with $LC$ deadline $D_i^L$, we can separately define $LC$ and $HC$ demand bound functions as follows (these functions were first defined for traditional non-mixed-criticality task systems~\cite{BMR90}).
\begin{align}	
dbf_i^L(t) & = \max \left \{ 0, \left ( \left \lfloor \frac{t - D_i^L}{T_i} \right \rfloor + 1  \right ) C_i^L \right \} \label{eqn:dbfl} \\
dbf_i^H(t) & = \max \left \{ 0, \left ( \left \lfloor \frac{t - D_i}{T_i} \right \rfloor + 1  \right ) C_i^H \right \} \label{eqn:dbfh}
\end{align}

Finally, to simplify presentation of equations, the following short-cut notation will be used in the paper.
\begin{equation*}
MOD(t,T_i) = t - \left \lfloor \frac{t}{T_i}  \right \rfloor T_i
\end{equation*}

%% file: schedulability.tex
\section{Schedulability Tests}
\label{sec:schedulability}

In this section we present a new schedulability test for mixed-criticality sporadic tasks, and show that it strictly dominates the only other known dbf-based test presented in ~\cite{EkYi12}.

\subsection{Existing schedulability test}
\label{sec:existing_test}

Ekberg and Yi~\cite{EkYi12} presented a dbf-based test by separately considering $LC$ and $HC$ system behaviors. In $LC$ behavior, each mixed-criticality task $\tau_i$ can be regarded as a traditional non-mixed-criticality task with deadline $D_i^L$ and worst-case execution time $C_i^L$. Therefore, EDF schedulability in $LC$ behavior can be checked using existing results as follows.

\begin{proposition}[From Theorem~1 in~\cite{BMR90}]
Task system $\tau$ is EDF schedulable in $LC$ system behavior if and only if,
\begin{equation*}
\forall 0 \leq t \leq t_{MAX}, \sum_{\tau_i \in \tau} dbf_i^L(t) \leq t, \mbox{ where}
\end{equation*}
$t_{MAX}$ is pseudo-polynomial in input size (defined in~\cite{BMR90}). 
\label{prop:LC_schedulability}
\end{proposition}

\begin{figure}
\centering
\includegraphics[width=0.8\linewidth]{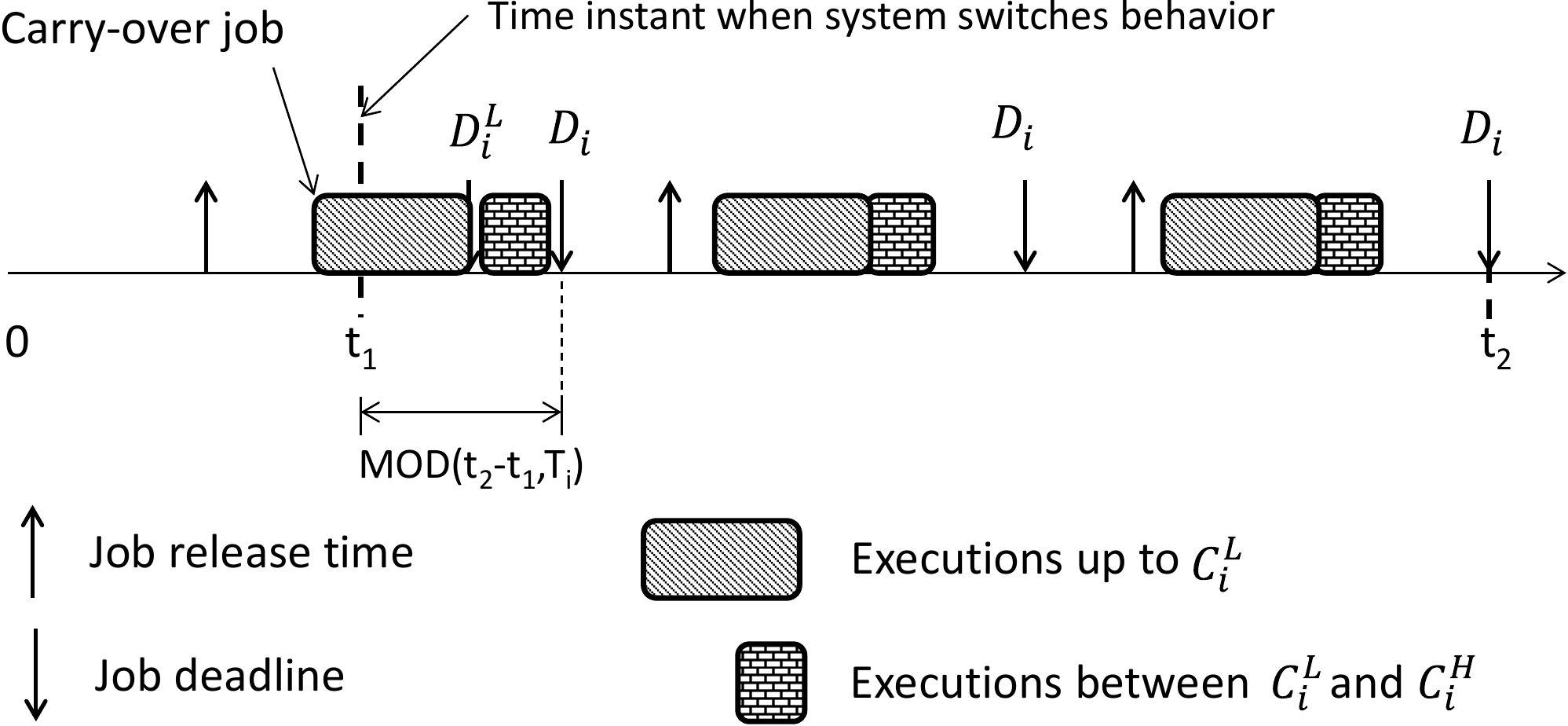}
\caption{Task execution pattern for maximum demand in $HC$ behavior}
\label{fig:existing_test}
\end{figure}

In $HC$ behavior, to determine the maximum demand of a $HC$ task $\tau_i$, we need to understand what is the contribution of the \emph{carry-over} job of this task. A carry-over job, as shown in Figure~\ref{fig:existing_test}, is a job of $\tau_i$ that is released before the time instant when the system switched from $LC$ to $HC$ behavior, and has a real deadline after this time instant. 
%Note if there are multiple such jobs, which can happen when task deadline is greater than minimum separation, then only the one that is released the earliest among them would be regarded as a carry-over job. This is because the remaining jobs will all be considered as executing only after the behavior switch so as to maximize the overall demand of the task. 
When computing demand in $HC$ behavior, we assume that all task deadlines in $LC$ behavior are always satisfied (verifiable using Proposition~\ref{prop:LC_schedulability}).

For an interval of length $t = t_2 - t_1$, $\tau_i$ generates maximum demand when some job of $\tau_i$ has a deadline at $t_2$, and all previous jobs of $\tau_i$ are released and execute as late as possible. This pattern is shown in Figure~\ref{fig:existing_test}. If $D_i^L$ of the carry-over job is before $t_1$, then it cannot contribute any execution to the interval, because it would have finished prior to the behavior switch. On the other hand, if $D_i^L$ is after $t_1$ as shown in the figure, then the remaining $LC$ execution of the carry-over job at $t_1$ is at most the interval length from $t_1$ to the tightened deadline. Otherwise, the job would not have met its deadline in $LC$ behavior if the switch did not happen at $t_1$. Therefore the total demand of this carry-over job is at most this remaining $LC$ execution plus $C_i^H - C_i^L$, which is the additional demand of the job in $HC$ behavior. The following proposition then presents the schedulability test for $HC$ behaviors.

\begin{proposition}[From~\cite{EkYi12}]
Task system $\tau$ is EDF schedulable in $HC$ system behavior if, $\forall t: 0 \leq t \leq t_{MAX}$,
\begin{equation*}
\sum_{\tau_i \in \mathcal{H}_{\tau}} dbf_i^H(t) + \sum_{\tau_i \in \mathcal{S}(t)} (C_i^H - C_i^L) + \sum_{\tau_i \in \mathcal{S}(t)} CO(t) \leq t.
\end{equation*}
Here $\mathcal{S}(t) = \left \{ \tau_i | \tau_i \in \mathcal{H}_{\tau} \mbox{ and } D_i > MOD(t,T_i) > D_i - D_i^L \right \}$ denotes tasks whose carry-over job will contribute to the demand in an interval of length $t$, and $CO(t) = \min \left \{ C_i^L, MOD(t,T_i) - (D_i - D_i^L) \right \}$ denotes the maximum carry-over executions that can be pending at the beginning of this interval.
\label{prop:HC_schedulability}
\end{proposition}

\subsection{New schedulability test}
\label{sec:new_test}

One of the main drawbacks of the test in Section~\ref{sec:existing_test} is that it cannot use the demand of tasks in $LC$ behavior to determine the remaining execution for carry-over jobs at the time of behavior switch. If the $LC$ demand of tasks is small, then many tasks would finish well before their deadlines. This means that many $HC$ carry-over jobs could also finish well before their tightened deadlines, and then the demand of these jobs in $HC$ behavior would decrease. The above test fails to use this dependency between $LC$ and $HC$ demands mainly because it bounds them individually. In this section we present a new dbf-based test that collectively bounds the $LC$ and $HC$ demands, and therefore is tighter than the above test.

Suppose there is a first deadline miss at some time instant $t_2$ in the schedule of task set $\tau$, and let $t_1$ denote a time instant when the system switches from $LC$ to $HC$ behavior such that $0 \leq t_1 \leq t_2$. When $t_1 = 0$ this represents a purely $HC$ behavior and when $t_1 = t_2$ this represents a purely $LC$ behavior, and therefore there is no loss of generality in this assumption. For purely $LC$ behaviors, that is when $t_1 = t_2 (= t \mbox{ say})$, we will continue to use the dbf-based test presented in Proposition~\ref{prop:LC_schedulability} because it is an exact test. In the remainder of this section we therefore only consider scenarios in which $t_1 < t_2$. Let $\mathcal{J}$ denote any minimal set of jobs of the task set $\tau$ whose schedule results in the deadline miss at $t_2$. Observe that by definition of minimality there are no idle instants in the schedule, because otherwise the schedule starting after the latest idle instant will also have a deadline miss at $t_2$ and the corresponding job set will be smaller.

We first derive an upper bound on the maximum demand that jobs of $LC$ tasks can have in this time interval through a series of lemmas as follows.

\begin{lemma}
No $LC$ job with deadline greater than $t_2$ can execute in the time interval $(0, t_1]$.
\label{lem:1}
\end{lemma}    
\begin{IEEEproof}
Suppose $LC$ jobs with deadline greater than $t_2$ execute in $(0, t_1]$, and let $t$ denote the latest time instant when any such job executes. Observe that no job with deadline smaller than or equal to $t_2$ is pending at this instant. Then the schedule resulting from jobs released at or after $t$ will also miss a deadline at $t_2$, contradicting the minimality of $\mathcal{J}$.
\end{IEEEproof}

\begin{figure}
\centering
\includegraphics[width=0.8\linewidth]{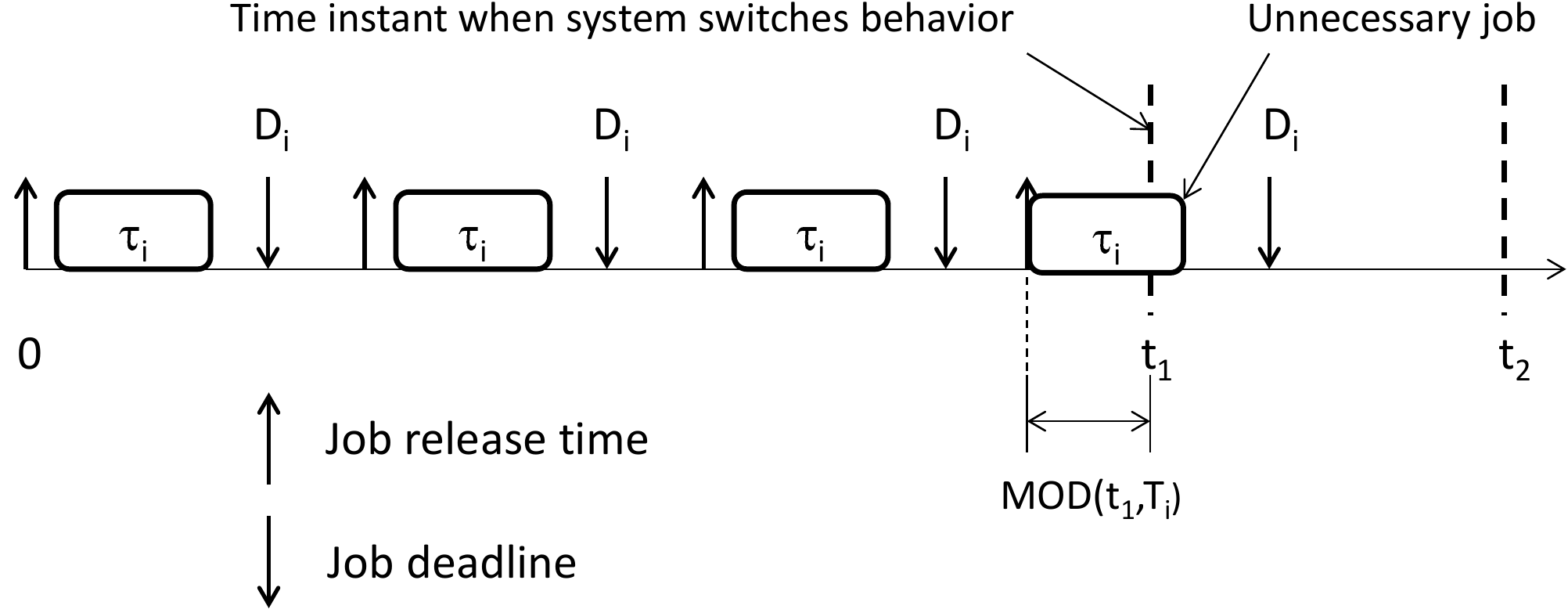}
\caption{Schedule for $LC$ task $\tau_i$ generating maximum demand}
\label{fig:lc_schedule}
\end{figure}

Thus only $LC$ jobs with deadline at most $t_2$ can execute in the interval. Among these, any job released at or after $t_1$ cannot execute because the system is already in $HC$ behavior. Therefore, apart from jobs that are both released and have deadline in the interval $(0, t_1]$, at most one job, called the \emph{unnecessary job}, can execute for each $LC$ task in $(0, t_1]$. This job has release time before $t_1$ and deadline in $(t_1, t_2]$. This job is unnecessary because its deadline is not required to be met, and if we had clairvoyance about the system switch at $t_1$, then we would never have executed it. This scenario is shown in Figure~\ref{fig:lc_schedule} for task $\tau_i$. To generate maximum demand in $(0, t_1]$, jobs are released as soon as possible. Further, the demand of unnecessary job is maximized if we assume that it executes continuously starting from its release time. Thus the following bound for demand of $LC$ task $\tau_i$ can be obtained.

\begin{lemma}
The maximum demand of $LC$ task $\tau_i$ in the time interval $(0, t_2]$ is given by,
\begin{equation*}
dbf_i(t_1, t_2) = dbf_i^L(t_1) + dbf_i^{UN}(t_1, t_2),
\end{equation*} 
where $dbf_i^L(t_1)$ is given by Equation~\eqref{eqn:dbfl}, and
\begin{align*}
& dbf_i^{UN}(t_1, t_2) = \\
& \begin{cases}
\min \left \{ C_i^L, MOD(t_1,T_i) \right \} & D_i^L > MOD(t_1,T_i) \\
& \mbox{ and } \left \lfloor \frac{t_1}{T_i} \right \rfloor T_i + D_i^L \leq t_2 \\
0 & Otherwise
\end{cases}
\end{align*}
\label{lem:lc_demand}
\end{lemma}

$dbf_i^{UN}(t_1, t_2)$ bounds the demand of unnecessary job of task $\tau_i$. Whenever $D_i^L > MOD(t_1,T_i)$, the deadline of the last job released in $(0, t_1]$ is after $t_1$. Condition $\left \lfloor \frac{t_1}{T_i} \right \rfloor T_i + D_i^L \leq t_2$ then ensures that this deadline is no later than $t_2$. 

To upper bound the demand of a $HC$ task $\tau_i$ we consider three different cases. The first case is when $t_2 - t_1 \leq D_i - D_i^L$. The following lemma asserts that in this case no job of $\tau_i$ can execute in $HC$ behavior.

\begin{lemma}
No job of $HC$ task $\tau_i$ satisfying inequality $t_2 - t_1 \leq D_i - D_i^L$ can execute in the interval $(t_1, t_2]$.
\end{lemma}
\begin{IEEEproof}
Suppose a job of $\tau_i$ has real deadline greater than $t_2$, but still executes in the interval $(t_1, t_2]$. Let $t$ denote the latest time instant when this job executes. Note that at $t$ no job with deadline at most $t_2$ is pending. Then the resulting schedule considering only jobs released at or after $t$ will also miss a deadline at time instant $t_2$. This contradicts the minimality of job set $\mathcal{J}$.

Thus, no job of $\tau_i$ with real deadline greater than $t_2$ can execute in the interval $(t_1, t_2]$. Suppose $\tau_i$ has a job with real deadline in the interval $(t_1, t_2]$. In this case, the tightened deadline of this job is at most $t_1$ because $t_2 - t_1 \leq D_i - D_i^L$. Then, the job would have already finished its execution before the behavior switch at $t_1$. This proves the lemma.
\end{IEEEproof}

Therefore, whenever $t_2 - t_1 \leq D_i - D_i^L$, $HC$ task $\tau_i$ essentially behaves like a $LC$ task, and its demand can be bounded using Lemma~\ref{lem:lc_demand}. An immediate corollary of this fact is that when $t_2 - t_1$ is smaller than the minimum difference between real and tightened deadlines of all $HC$ tasks, then no $HC$ job can execute in the interval $(t_1, t_2]$, and therefore the deadline miss scenario at $t_2$ is not feasible.

\begin{corollary}
For a $HC$ deadline miss to occur at time $t_2$ when the system switches behavior at time $t_1 (\leq t_2)$, it must be the case that $t_2 - t_1 > \min_{\tau_i \in \mathcal{H}_{\tau}} \{ D_i - D_i^L \}$.
\label{cor:1}
\end{corollary}

\begin{figure}
\centering
\includegraphics[width=0.8\linewidth]{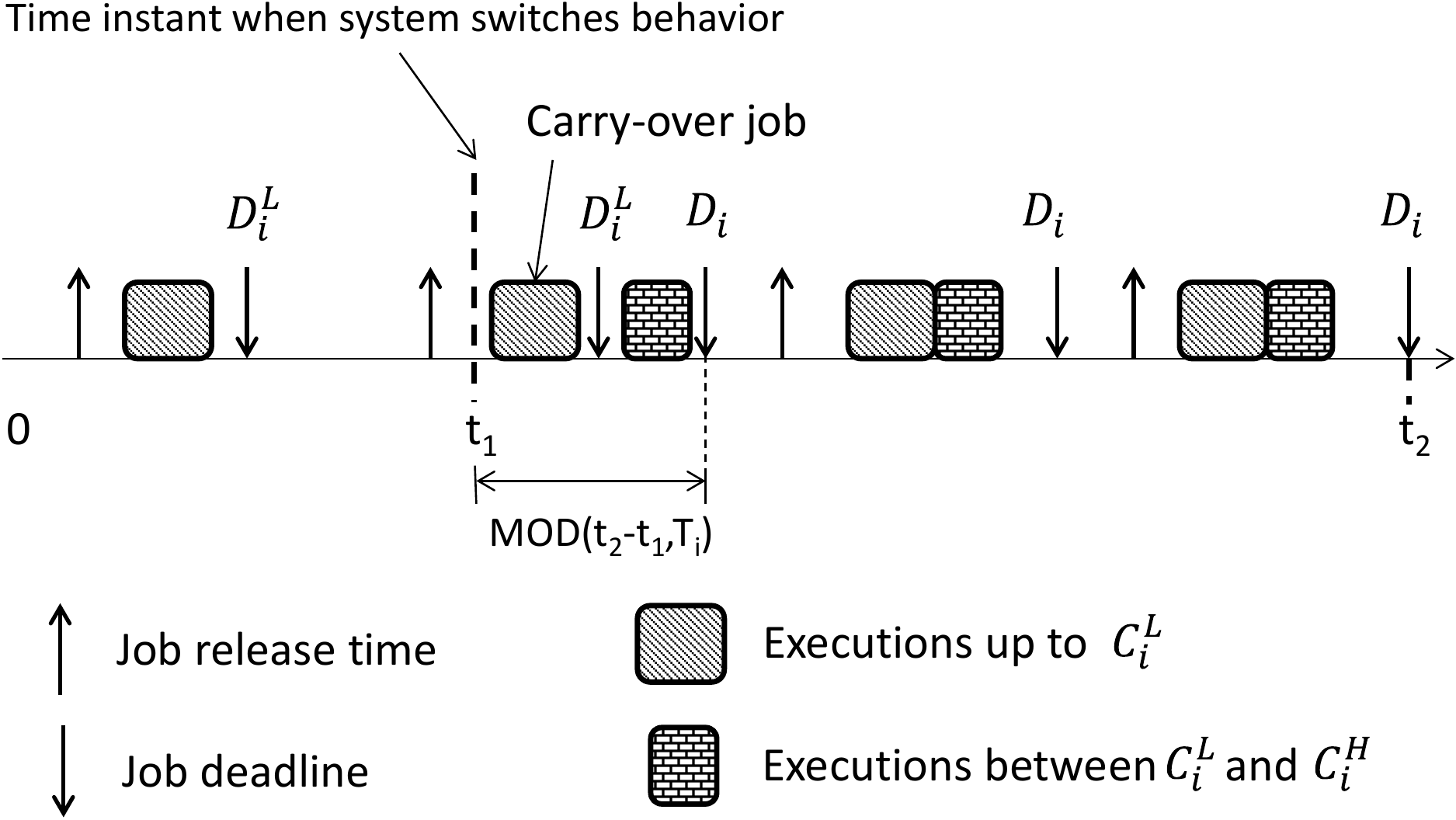}
\caption{Schedule of $HC$ task $\tau_i$ when carry-over job executes after $t_1$}
\label{fig:hc_schedule_case2}
\end{figure}

The second case is when $t_2 - t_1 > D_i - D_i^L$ and the carry-over job can contribute demand in $HC$ behavior. This scenario is depicted in Figure~\ref{fig:hc_schedule_case2}. As shown, the carry-over job can contribute demand in $HC$ behavior when its tightened deadline is greater than $t_1$. Further, to maximize overall demand of task $\tau_i$ it is necessary to assume that this carry-over job does not finish before the behavior switch at $t_1$. In this case, it will contribute an additional demand of $C_i^H - C_i^L$. 

The overall demand of task $\tau_i$ in the interval $(0, t_2]$ is maximized when the real deadline of a job of this task coincides with time instant $t_2$, all preceding jobs are released as late as possible, and the carry-over job executes as late as possible. This pattern is shown in Figure~\ref{fig:hc_schedule_case2}. Suppose we shift the releases to the right by some amount smaller than $T_i$. Then the demand from the last $HC$ job would decrease by $C_i^H$ because its deadline is no longer in the interval. The demand from the carry-over job may increase by at most $C_i^H - C_i^L$ if its tightened deadline moves to a time instant greater than $t_1$ after the shift. The demand from $LC$ jobs may increase by at most $C_i^L$ if an additional $LC$ job can now be accommodated in the interval. Therefore, the total increase in demand is at most $C_i^H$ which is no larger than the total decrease in demand. Therefore this task release and execution pattern maximizes the demand of task $\tau_i$ in the interval $(0, t_2]$. The following lemma derives a bound for the demand of task $\tau_i$ in this case.

\begin{lemma}
When $D_i - D_i^L < MOD(t_2-t_1,T_i) < D_i$ and $\left \lfloor \frac{t_2 - t_1}{T_i} \right  \rfloor T_i + D_i \leq t_2$, the demand of $HC$ task $\tau_i$ is given by,
{\small \begin{align*}
& dbf_i(t_1, t_2) = dbf_i^L(t_1, t_2) + dbf_i^H(t_1, t_2) + dbf_i^{CO}(t_1, t_2), \\
& \\
& \mbox{where } \\
& dbf_i^L(t_1, t_2) = \max \left \{ 0, \left ( \left \lfloor \frac{t_2 - D_i}{T_i} \right \rfloor - \left \lfloor \frac{t_2 - t_1 - D_i}{T_i} \right \rfloor - 1 \right ) \right \} C_i^L, \\
& dbf_i^H(t_1, t_2) = \max \left \{ 0, \left ( \left \lfloor \frac{t_2 - t_1 - D_i}{T_i} \right \rfloor + 1 \right ) C_i^H \right \}, \mbox{ and} \\
& dbf_i^{CO}(t_1, t_2) = C_i^H.
\end{align*}}
\label{lem:hc_2}
\end{lemma}
\begin{IEEEproof}
Condition $D_i - D_i^L < MOD(t_2-t_1,T_i) < D_i$ checks for the existence of a carry-over job that is released before $t_1$ and has tightened deadline in the interval $(t_1, t_2]$. Condition $\left \lfloor \frac{t_2 - t_1}{T_i} \right  \rfloor T_i + D_i \leq t_2$ on the other hand checks whether the interval $(0, t_2]$ is large enough to accommodate this carry-over job. Hence, if these two 
conditions are met, there exists a carry-over job that will contribute demand in the interval $(t_1, t_2]$. The total demand of this carry-over job, denoted as $dbf_i^{CO}(t_1, t_2)$, is $C_i^H$ because it completes execution in $HC$ behavior.

The total demand of all the $HC$ jobs in the interval $(t_1, t_2]$, excluding the carry-over job, is given by the traditional definition of demand bound function for non-mixed-criticality systems (see Equation~\eqref{eqn:dbfh}).

The total number of jobs in the interval $(0, t_2]$ is $\left \lfloor \frac{t_2 - D_i}{T_i} \right \rfloor+1$. Of these,  $\left \lfloor \frac{t_2 - t_1 - D_i}{T_i} \right \rfloor+1$ jobs execute only in the interval $(t_1, t_2]$ and their demand is considered in $dbf_i^H(t_1, t_2)$. Additionally, the demand of carry-over job is also accounted for. Therefore, the number of remaining jobs that only contribute demand in the interval $(0, t_1]$ is $\left \lfloor \frac{t_2 - D_i}{T_i} \right \rfloor + 1 - \left \lfloor \frac{t_2 - t_1 - D_i}{T_i} \right \rfloor - 1 - 1$. Since each such job can execute for a maximum time of $C_i^L$, the total demand in this interval, denoted as  $dbf_i^L(t_1, t_2)$, is bounded by $\left ( \left \lfloor \frac{t_2 - D_i}{T_i} \right \rfloor - \left \lfloor \frac{t_2 - t_1 - D_i}{T_i} \right \rfloor - 1 \right ) C_i^L$.
\end{IEEEproof}

\begin{figure}
\centering
\includegraphics[width=0.8\linewidth]{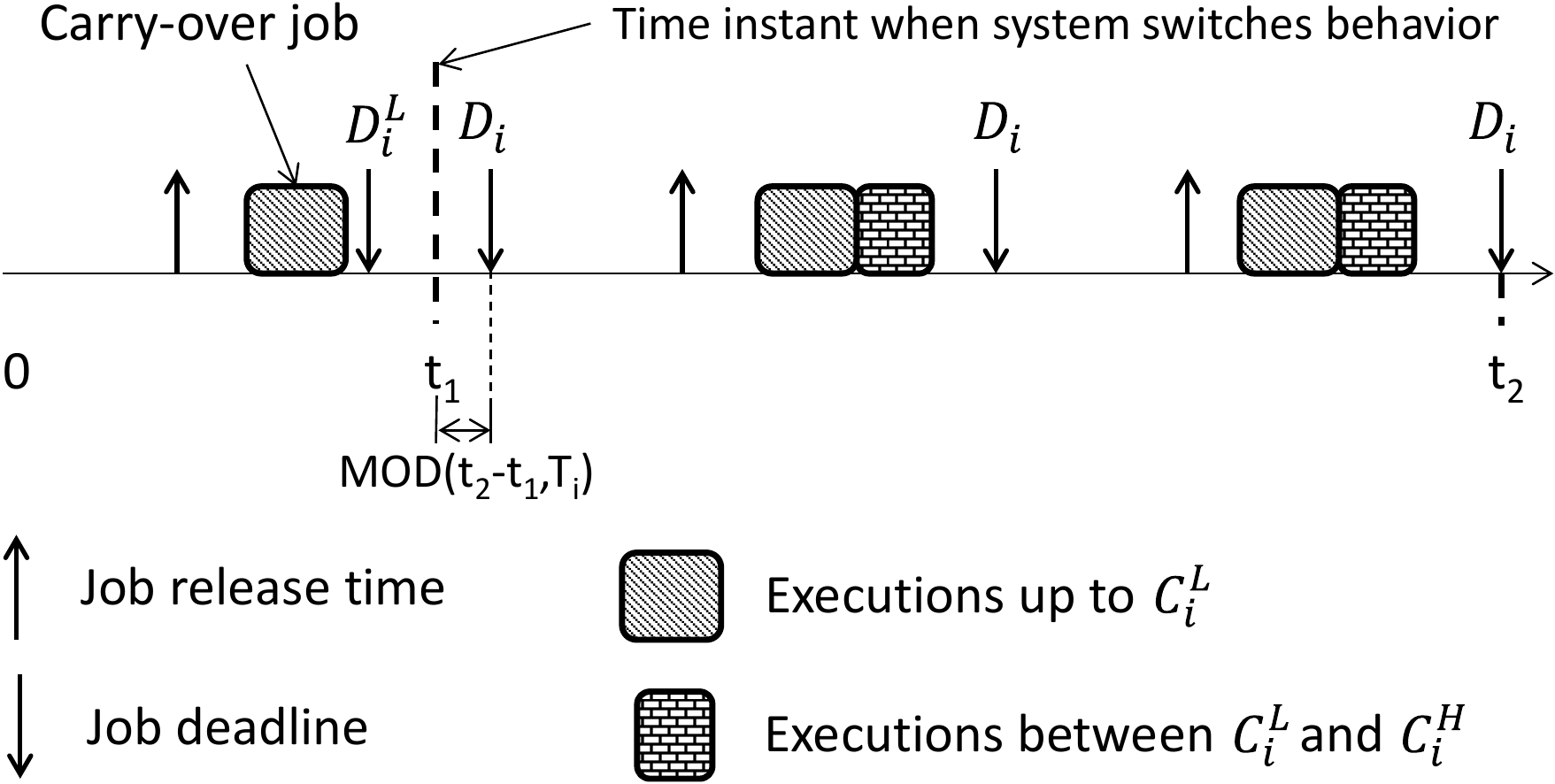}
\caption{Schedule of $HC$ task $\tau_i$ when carry-over job finishes by $t_1$}
\label{fig:hc_schedule_case3}
\end{figure}

The third case is when $t_2 - t_1 > D_i - D_i^L$ and the carry-over job cannot contribute demand in $HC$ behavior. This scenario is depicted in Figure~\ref{fig:hc_schedule_case3}. The tightened deadline of the carry-over job is no greater than $t_1$ and hence the job would finish before the system switches to $HC$ behavior. The total demand of all the jobs of $\tau_i$ in this case is similar to the previous case, except that the carry-over job only contributes $C_i^L$ instead of $C_i^H$. We record this result in the following lemma.

\begin{lemma}
When conditions in Lemmas~\ref{lem:lc_demand} and~\ref{lem:hc_2} are not satisfied, then the demand of $HC$ task $\tau_i$ is given by,
\begin{equation*}
dbf_i(t_1, t_2) = dbf_i^L(t_1, t_2) + dbf_i^H(t_1, t_2) + dbf_i^{CO}(t_1, t_2),
\end{equation*}
where $dbf_i^{CO}(t_1, t_2) = C_i^L$, and $dbf_i^L(t_1, t_2)$ and $dbf_i^H(t_1, t_2)$ are given in Lemma~\ref{lem:hc_2}.
\label{lem:hc_3}
\end{lemma}

Thus, whenever there is a deadline miss at some instant $t_2$ and the system switches behavior at some instant $t_1$ ($< t_2$), an upper bound on the total demand of any minimal job set that leads to this deadline miss is as discussed above. Since there is a deadline miss at $t_2$ and there are no idle instants in the interval $(0, t_2]$, the total demand of this minimal job set (and therefore its upper bound) must exceed $t_2$. Using the contrapositive of this statement and Corollary~\ref{cor:1}, we get the following schedulability test.

\begin{theorem}
Task system $\tau$ is EDF schedulable in $HC$ system behavior if,
\begin{align*}
& \forall t_2:  0 \leq t_2 \leq t_{MAX},  \forall t_1: 0 \leq t_1 < t_2 - \min_{\tau_i \in \mathcal{H}_{\tau}} \left \{ D_i - D_i^L \right \}, \\ 
& \sum_{\tau_i \in \tau} dbf_i(t_1, t_2) \leq t_2.
\end{align*}
Here $dbf_i(t_1, t_2)$ is given by Lemma~\ref{lem:lc_demand} when $\tau_i$ is a $LC$ task or a $HC$ task in case~1, by Lemma~\ref{lem:hc_2} if it is a $HC$ task in case~2, and by~\ref{lem:hc_3} if it is a $HC$ task in case~3.
\label{thm:new_test}
\end{theorem}

\subsection{Improved schedulability test}
\label{sec:improved_test}

The schedulability test presented in Theorem~\ref{thm:new_test} collectively bounds the demand of $HC$ and $LC$ tasks. However, the demand bound in $LC$ behavior (time interval $(0, t_1]$) is pessimistic, and in this section we tighten it.

\begin{figure}
\centering
\includegraphics[width=0.8\linewidth]{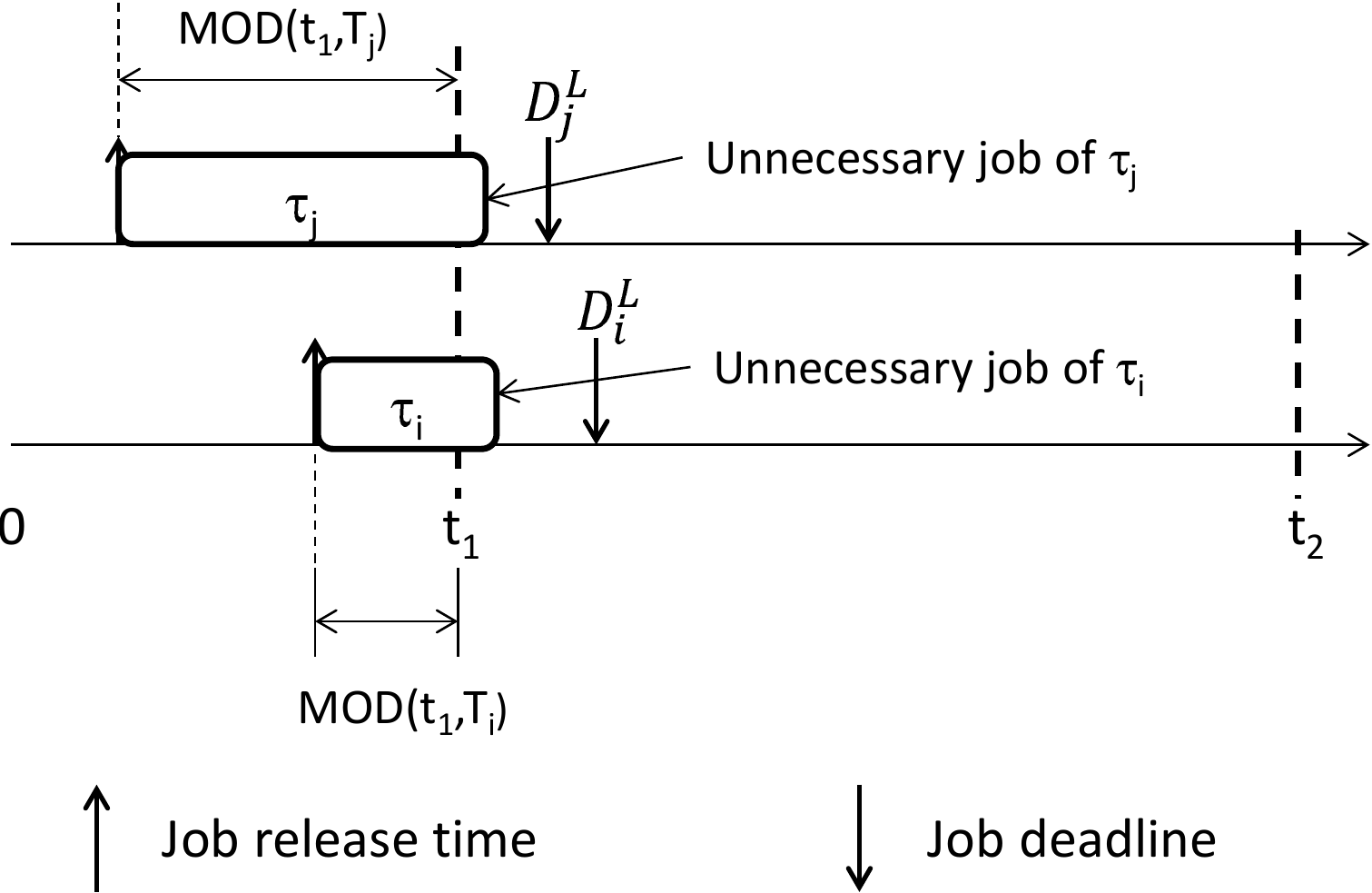}
\caption{Pessimistic demand bound for unnecessary jobs}
\label{fig:unnecessary_pessimism}
\end{figure}

An unnecessary job, as shown in Figure~\ref{fig:lc_schedule}, is a job that is released before $t_1$, has a tightened deadline in the interval $(t_1, t_2]$, and if the job belongs to a $HC$ task then its real deadline is after $t_2$. The contribution of each such job to the total demand in Theorem~\ref{thm:new_test} is given by $dbf_i^{UN}(t_1,t_2)  = \min \left \{ C_i^L, MOD(t_1,T_i) \right \}$ (Lemma~\ref{lem:lc_demand}). That is, the job is assumed to contribute either $C_i^L$ or the interval length between its release time and $t_1$, whichever is smaller. This is however very pessimistic for some cases, such as the one shown in Figure~\ref{fig:unnecessary_pessimism}. Here tasks $\tau_i$ and $\tau_j$ both have unnecessary jobs, and therefore their total demand bound as given by Theorem~\ref{thm:new_test} would be $MOD(t_1,T_i) + MOD(t_1,T_j)$. But this is not possible because the two jobs cannot run in parallel on an uniprocessor and cannot execute after $t_1$. In fact, the maximum demand that these two jobs can collectively generate is bounded by $\max \{ D_i^L, D_j^L\}$ and this can be explained as follows. By definition, any unnecessary job of $\tau_i$ or $\tau_j$ must be released in the time interval $[t_1 - \max \{ D_i^L, D_j^L\}, t_1)$. Then, in the worst case, these unnecessary jobs continuously execute from this earliest release time until $t_1$, and therefore their total demand cannot exceed $\max \{ D_i^L, D_j^L\}$. This argument can be easily generalized to an arbitrary number of tasks, and the resulting bound is recorded in the following lemma.

\begin{lemma}
The total demand of all the unnecessary jobs collectively is given by,
\begin{align*}
& dbf_{UN}(t_1, t_2) = \\
& \min \left \{ \max_{\stackrel{\tau_i \in \mathcal{L}_{\tau} \mbox{ {\scriptsize or}}}{\tau_i \in \mathcal{H}_{\tau} \mbox{ {\scriptsize and case~1}}}} \left \{ D_i^L \right \},  \sum_{\stackrel{\tau_i \in \mathcal{L}_{\tau} \mbox{ {\scriptsize or}}}{\tau_i \in \mathcal{H}_{\tau} \mbox{ {\scriptsize and case~1}}}} dbf_i^{UN}(t_1, t_2) \right \},
\end{align*}
where $dbf_i^{UN}(t_1, t_2)$ is defined in Lemma~\ref{lem:lc_demand}.
\label{lem:unnecessary_bound}
\end{lemma}    

Another source of pessimism is the $LC$ demand in the time interval $(0, t_1]$. Since there is no deadline miss in the interval $(0, t_1]$, the total demand in this interval cannot exceed $t_1$. We now look at the total demand for the interval $(0, t_2]$ in Theorem~\ref{thm:new_test} and Lemma~\ref{lem:unnecessary_bound}, and identify the minimum of this total demand that belongs to the interval $(0, t_1]$. The following lemmas determine this minimum demand for different tasks.

\begin{lemma}
Out of the total demand in interval $(0, t_2]$ for $LC$ tasks and $HC$ tasks in case~1 (Lemmas~\ref{lem:lc_demand} and~\ref{lem:unnecessary_bound}), the minimum demand in interval $(0, t_1]$ is given by,
\begin{equation*}
dbf_{L_1}(t_1, t_2) =  dbf_{UN}(t_1, t_2) + \sum_{\stackrel{\tau_i \in \mathcal{L}_{\tau} \mbox{ {\scriptsize or}}}{\tau_i \in \mathcal{H}_{\tau} \mbox{ {\scriptsize and case~1}}}} dbf_i^L(t_1),
\end{equation*}
where $dbf_{UN}(t_1, t_2)$ is given by Lemma~\ref{lem:unnecessary_bound} and $dbf_i^L(t_1)$ is given by Equation~\eqref{eqn:dbfl}.
\label{lem:lc_min_demand}
\end{lemma}
\begin{IEEEproof}
All the unnecessary jobs can only generate demand in $(0, t_1]$, because they will be ignored after $t_1$. The remaining demand of all $LC$ tasks and $HC$ tasks in case~1 must also belong to $(0, t_1]$ because they have their release and deadline in this interval. Therefore, for both these cases, their entire demand belongs to the interval $(0, t_1]$.
\end{IEEEproof}

\begin{lemma}
Out of the total demand in interval $(0, t_2]$ for $HC$ tasks in case~2 (Lemma~\ref{lem:hc_2}), the minimum demand in interval $(0, t_1]$ is given by,
\begin{equation*}
dbf_{L_2}(t_1, t_2) =  \sum_{\stackrel{\tau_i \in \mathcal{H}_{\tau}}{\mbox{{\scriptsize and case~2}}}} \left ( dbf_i^L(t_1, t_2) + C_i^L - CO(t_2-t_1) \right ),
\end{equation*}
where $dbf_i^L(t_1, t_2)$ is given by Lemma~\ref{lem:hc_2}, and $CO(t_2-t_1)$ is defined in Proposition~\ref{prop:HC_schedulability}.
\label{lem:hc2_min_demand}
\end{lemma}
\begin{IEEEproof}
For a $HC$ task in case~2, the carry-over job can potentially generate demand in both the intervals $(0, t_1]$ and $(t_1, t_2]$. The maximum demand this carry-over job can generate in the interval $(t_1, t_2]$ is $CO(t_2-t_1) + C_i^H - C_i^L$ (see Figure~\ref{fig:existing_test}). $CO(t_2-t_1) = \min \left \{ C_i^L, MOD(t_2-t_1,T_i) - (D_i - D_i^L) \right \}$ is the maximum executions that can be pending at time $t_1$, because otherwise the job would not meet its deadline if the system did not switch to $HC$ behavior. $C_i^H - C_i^L$ is the additional demand generated by this job in $HC$ behavior. Therefore, the minimum demand of this job in the interval $(0, t_1]$ is $C_i^L - CO(t_2-t_1)$. Additionally, the entire demand of $dbf_i^L(t_1, t_2)$ is generated by jobs that are both released and have their deadlines in the interval $(0, t_1]$, and therefore it belongs to the same interval.  
\end{IEEEproof}

\begin{lemma}
Out of the total demand in interval $(0, t_2]$ for $HC$ tasks in case~3 (Lemma~\ref{lem:hc_3}), the minimum demand in interval $(0, t_1]$ is given by,
\begin{equation*}
dbf_{L_3}(t_1, t_2) =  \sum_{\stackrel{\tau_i \in \mathcal{H}_{\tau}}{\mbox{{\scriptsize and case~3}}}} \left ( dbf_i^L(t_1, t_2) + dbf_i^{CO}(t_1, t_2) \right ),
\end{equation*}
where $dbf_i^L(t_1, t_2)$ and $dbf_i^{CO}(t_1, t_2)$ are given by Lemma~\ref{lem:hc_3}.
\label{lem:hc3_min_demand}
\end{lemma}
\begin{IEEEproof}
For a $HC$ task in case~3, since the carry-over job does not generate demand in $(t_1, t_2]$, its entire demand belongs to the interval $(0, t_1]$. Additionally, similar to case~2 above, the entire demand of $dbf_i^L(t_1, t_2)$ also belongs to this interval.
\end{IEEEproof}

The following theorem then presents an improved EDF schedulability test for $HC$ behaviors. 

\begin{theorem}
Task system $\tau$ is EDF schedulable in $HC$ system behavior if,
\begin{align*}
& \forall t_2:  0 \leq t_2 \leq t_{MAX},  \forall t_1: 0 \leq t_1 < t_2 - \min_{\tau_i \in \mathcal{H}_{\tau}} \left \{ D_i - D_i^L \right \}, \\ 
& \min \left \{ t_1, \sum_{j=1}^3 dbf_{L_j}(t_1, t_2) \right \} + \sum_{\stackrel{\tau_i \in \mathcal{H}_{\tau}}{\mbox{{\scriptsize and case~2 or~3}}}} dbf_i^H(t_1, t_2) \\
& + \sum_{\stackrel{\tau_i \in \mathcal{H}_{\tau}}{\mbox{{\scriptsize and case~2}}}} \left ( CO(t_2 - t_1) + C_i^H - C_i^L \right ) \leq t_2.
\end{align*}
Here $dbf_i^H(t_1, t_2)$ is given by Lemma~\ref{lem:hc_2}, $dbf_{L_1}(t_1, t_2)$ by Lemma~\ref{lem:lc_min_demand}, $dbf_{L_2}(t_1, t_2)$ by Lemma~\ref{lem:hc2_min_demand}, and $dbf_{L_3}(t_1, t_2)$ by Lemma~\ref{lem:hc3_min_demand}. Also, $CO(t_2-t_1)$ is defined in Proposition~\ref{prop:HC_schedulability}. 
\label{thm:improved_test}
\end{theorem}
\begin{IEEEproof}
From Lemmas~\ref{lem:lc_min_demand},~\ref{lem:hc2_min_demand}, and~\ref{lem:hc3_min_demand}, we know that the total minimum demand in the interval $(0, t_1]$ is $dbf_{L_1}(t_1, t_2)  + dbf_{L_2}(t_1, t_2) + dbf_{L_3}(t_1, t_2)$, and this demand cannot exceed $t_1$ because there is no deadline miss in $(0, t_1]$.

The remaining demand in interval $(0, t_2]$ is the $HC$ demand $dbf_i^H(t_1, t_2)$ for $HC$ tasks in cases~2 and~3, as well as the carry-over demand in interval $(t_1, t_2]$ for $HC$ tasks in case~2. This carry-over demand for a task $\tau_i$ is $C_i^H - C_i^L$ representing the additional execution required in $HC$ behavior, and $CO(t_2 - t_1)$ representing the maximum remaining execution for the carry-over job at time instant $t_1$.
\end{IEEEproof}

\subsection{Test properties}
\label{sec:properties}

In this section we show that the test presented in Theorem~\ref{thm:improved_test} strictly dominates the one presented in Proposition~\ref{prop:HC_schedulability}. We also briefly discuss the test complexity.

For $LC$ behaviors, both the existing test as well as the new test use the same condition presented in Proposition~\ref{prop:LC_schedulability}. For $HC$ behaviors, the following theorem shows that the new test dominates the existing test. 

\begin{theorem}
If a task system $\tau$ is EDF schedulable in $HC$ behaviors based on Proposition~\ref{prop:HC_schedulability}, then it is also EDF schedulable in $HC$ behaviors based on Theorem~\ref{thm:improved_test}.
\end{theorem}
\begin{IEEEproof}
Consider some time instant $t$ for which the condition in Proposition~\ref{prop:HC_schedulability} holds. Now we show that the condition in Theorem~\ref{thm:improved_test} also holds for all $t_1$ and $t_2$ such that $t_2 - t_1 = t$.
\begin{align*}
& \mbox{LHS of condition in Theorem~\ref{thm:improved_test}} \\
\leq & t_1 + \sum_{\stackrel{\tau_i \in \mathcal{H}_{\tau} \mbox{ {\scriptsize and}}}{\mbox{{\scriptsize case~2 or~3}}}} dbf_i^H(t_1, t_2) \\
& + \sum_{\stackrel{\tau_i \in \mathcal{H}_{\tau}}{\mbox{{\scriptsize and case~2}}}} \left ( CO(t_2 - t_1) + C_i^H - C_i^L \right ) \\
& \\
& \mbox{($dbf_i^H(t_1, t_2) = 0$ for $HC$ tasks in case~1 $\Rightarrow$)} \\
& \\
= & t_1 + \sum_{\tau_i \in \mathcal{H}_{\tau}} dbf_i^H(t_1, t_2) \\
& + \sum_{\stackrel{\tau_i \in \mathcal{H}_{\tau}}{\mbox{{\scriptsize and case~2}}}} \left ( CO(t_2 - t_1) + C_i^H - C_i^L \right ) \\
\end{align*}
\begin{align*}
& \mbox{(Using $dbf_i^H(t_1, t_2) = dbf_i^H(t_2-t_1)$ from Lemma~\ref{lem:hc_2}, and} \\
& \mbox{$\tau_i \in \mathcal{S}(t_2-t_1)$ implies $\tau_i$ in case~2 from Proposition~\ref{prop:HC_schedulability})} \\
& \\
= & t_1 + \sum_{\tau_i \in \mathcal{H}_{\tau}} dbf_i^H(t_2-t_1) \\
& + \sum_{\tau_i \in \mathcal{S}(t_2-t_1)} \left ( CO(t_2 - t_1) + C_i^H - C_i^L \right ) \\
& \\
& \mbox{(Proposition~\ref{prop:HC_schedulability} is satisfied for $t=t_2-t_1$ $\Rightarrow$)} \\
& \\
\leq & t_1 +  t_2 - t_1 = t_2.
\end{align*}

This shows that whenever the condition in Proposition~\ref{prop:HC_schedulability} holds for some time $t$, conditions in Theorem~\ref{thm:improved_test} for time instants $t_1$ and $t_2$ satisfying $t_2 - t_1 = t$ also hold. Therefore, when Proposition~\ref{prop:HC_schedulability} holds for all time instants $t: 0 \leq t \leq t_{MAX}$, Theorem~\ref{thm:improved_test} also holds for all time instants $t_1$ and $t_2$ such that $0 \leq t_2 \leq t_{MAX}$ and $t_1 < t_2$.
\end{IEEEproof}

While the above theorem proves dominance of the new test, the following example shows that this dominance is strict.
\begin{example}
Consider task system comprising of two tasks $\tau_1 = \{ 6, 4, HC, \{ 1, 2 \} \}$ and $\tau_2 = \{ 7, 5, LC, \{ 1, 1 \} \}$. Recall from Section~\ref{sec:model} that a dual-criticality sporadic task is specified as $\tau_i = (T_i, D_i, L_i, \mathcal{C}_i)$, where $T_i$ denotes minimum separation between successive job releases, $D_i$ denotes deadline, $L_i$ denotes criticality level, and $\mathcal{C}_i = \{ C_i^L, C_i^H \}$ is a list of WCET values with $C_i^L$ denoting the low-criticality (or $LC$) WCET and $C_i^H$ denoting the high-criticality (or $HC$) WCET.

Let us assume that the tightened deadlines of these tasks are the same as their real deadlines, that is, $D_1^L = D_1 = 4$ and $D_2^L = D_2 = 5$. It can be easily verified that Proposition~\ref{prop:HC_schedulability} fails for $t = 1$, because it computes the demand of task $\tau_1$ in this interval as $2$ time units. On the other hand, Theorem~\ref{thm:improved_test} succeeds for this task system, and in fact, this task system is easily schedulable even if we reserve $2$ time units for task $\tau_1$ at all times.
\end{example}

\textbf{Test complexity.} For a given set of time instants $t_1$ and $t_2$, the condition in Theorem~\ref{thm:improved_test} and Proposition~\ref{prop:LC_schedulability} can be evaluated in time proportional to the number of tasks in $\tau$. The bound on time instants ($t_{MAX}$) is a pseudo-polynomial in the size of the input (see~\cite{BMR90}). Therefore, the overall complexity of the proposed test is also pseudo-polynomial, although there is a quadratic increase in the number of conditions to be evaluated when compared to the existing test. 

%% file: simulations.tex
\section{Deadline tightening strategy}
\label{sec:strategy_and_simulations}

In this section we present a new deadline tightening strategy that uses the schedulability test presented in Theorem~\ref{thm:improved_test} and evaluate its performance through extensive simulations.

\subsection{Strategy}
\label{sec:strategy}

Algorithm~\ref{alg:ECDF} presents our deadline tightening strategy called Earliest Carry-over Deadline First (ECDF). It first checks for satisfaction of the $LC$ schedulability test in Lines~\ref{line:1}--\ref{line:2} (Proposition~\ref{prop:LC_schedulability}), and then for satisfaction of the $HC$ schedulability test in Lines~\ref{line:3}--\ref{line:4} (Theorem~\ref{thm:improved_test}). If the test for $HC$ behaviors failed for some instants $t_1$ and $t_2$, then we identify an \emph{appropriate} candidate among all the $HC$ tasks and tighten its $LC$ deadline by $1$ (Line~\ref{line:5}). If the test for $LC$ behaviors failed for some time instant after some deadline was tightened, then we backtrack and increment the deadline by $1$ (Line~\ref{line:68}). The list of candidate $HC$ tasks is continuously updated in each step, so that a task is removed from this list if its $LC$ deadline cannot be reduced anymore as in Line~\ref{line:7}, or if tightening its deadline resulted in a failed test for $LC$ behaviors as in Line~\ref{line:68}. Thus, in each step, either some task deadline is reduced by one or some task is removed from the list of candidates, and therefore the while loop in the algorithm is guaranteed to terminate.

\begin{algorithm}
\caption{ECDF: Earliest Carry-over Deadline First}
\label{alg:ECDF}
\begin{algorithmic}[1]
\State $i \leftarrow \perp$ and candidates $\leftarrow \left \{ \tau_i | \tau_i \in \mathcal{H}_{\tau} \right \}$.
\While{True}
\State feasible $\leftarrow$ true.
\For{$t=0 \ldots t_{MAX}$}\label{line:1}
\If{Proposition~\ref{prop:LC_schedulability} fails}
\State If $i = \perp$, return failure.
\State $D_i^L = D_i^L + 1$ and remove $\tau_i$ in candidates.\label{line:68}
\State $i \leftarrow \perp$ and break.
\EndIf
\EndFor\label{line:2}
\For{$t_2 = 0 \ldots t_{MAX}$ and $t_1 = 0 \ldots t_2 - \min_{\tau_i \in \mathcal{H}_{\tau}} \{ D_i - D_i^L \} - 1$} \label{line:3}
\If{Theorem~\ref{thm:improved_test} fails}
\State If $t_1 = 0$ or candidates $= \Phi$, return failure.
\State i = FINDCANDIDATE(candidates, $t_1$, $t_2$).
\State $D_i^L = D_i^L - 1$.\label{line:5}
\State If $D_i^L - 1 < C_i^L$, remove $\tau_i$ in candidates.\label{line:7}
\State feasible $\leftarrow$ false and break.
\EndIf
\EndFor \label{line:4}
\State If feasible is true, return success.
\EndWhile
\Function{findCandidate}{candidates, $t_1$, $t_2$}
\State Let $DEM$ denote the excess demand at time instant $t_2$ (LHS of Theorem~\ref{thm:improved_test} - $t_2$).
\State $result \leftarrow \perp$, $DIFF = 0$, $DEC = \infty$.
\For{Each task $\tau_i$ in candidates}
%\If{$\tau_i$ belongs to case~2 (Lemma~\ref{lem:hc_2} in Section~\ref{sec:new_test}) and $C_i^H - C_i^L \geq EXCESS$} \label{line:13}
\If{$\tau_i$ in case~2 and $C_i^H - C_i^L \geq DEM$} \label{line:13}
\If{{\small $MOD(t_2-t_1, T_i) - (D_i-D_i^L) < DEC$}} \label{line:10}
\State $DEC \leftarrow MOD(t_2-t_1, T_i) - (D_i-D_i^L)$.
\State $result \leftarrow i$ and $DIFF \leftarrow C_i^H - C_i^L$.
\ElsIf{$MOD(t_2-t_1, T_i) - (D_i-D_i^L) = DEC$ and $C_i^H - C_i^L > DIFF$} \label{line:11}
\State $result \leftarrow i$ and $DIFF \leftarrow C_i^H - C_i^L$.
\EndIf \label{line:12}
\EndIf
\EndFor 
\State Return result.
\EndFunction
\end{algorithmic}
\end{algorithm}

Function FINDCANDIDATE presented in Algorithm~\ref{alg:ECDF} identifies the appropriate $HC$ task whose deadline must be tightened from among a list of candidates. In Section~\ref{sec:new_test} we split the set of $HC$ tasks into three cases for each time instant pair $t_1$ and $t_2$. Tasks in case~1 do not generate any $HC$ demand and their demand bound is given by Lemma~\ref{lem:lc_demand}. It is easy to see from this lemma that if the deadline $D_i^L$ of this task is tightened, then it continues to be in case~1 for the same time instant pair, and further its demand bound may only increase. Therefore, these tasks are not good candidates for deadline tightening. Now consider a $HC$ task in case~3 whose demand bound is given in Lemma~\ref{lem:hc_3}, that is, a task whose carry-over job does not generate any demand in the interval $(t_1, t_2]$. Therefore, the $LC$ deadline $D_i^L$ of this carry-over job is no later than $t_1$, and tightening this deadline will not change the demand of the carry-over job. It is also easy to see that tightening $D_i^L$ will not change the demand of any other job of this task either. Thus, tasks in case~3 are also not good candidates for tightening the $LC$ deadlines.

Now consider a $HC$ task $\tau_i$ in case~2 whose demand bound is given by Lemma~\ref{lem:hc_2}. If we tighten its $LC$ deadline $D_i^L$ so that $D_i - D_i^L > MOD(t_2-t_1, T_i)$, then the carry-over job can no longer contribute demand in the interval $(t_1, t_2]$ and therefore its demand would decrease by $C_i^H - C_i^L$. We use this property to identify the appropriate candidate in function FINDCANDIDATE. From among all the tasks in case~2, we choose the task that requires the smallest change in $LC$ deadline to cause this demand reduction (Line~\ref{line:10}), and hence the name Earliest Carry-over Deadline First. If there is a tie, then we break the tie using the largest reduction in demand first strategy as shown in Lines~\ref{line:11}--\ref{line:12} (largest value for $C_i^H - C_i^L$). Further, as a small optimization, we only consider those tasks in case~2 whose demand reduction would result in the failed schedulability test for time instant pair $t_1$ and $t_2$ now being satisfied (check in Line~\ref{line:13}). Note that this proposed strategy is almost identical to the strategy earlier proposed by Ekberg and Yi~\cite{EkYi12}, but for two crucial exceptions. One is that we use the improved test, and the second is our strategy for identifying an appropriate candidate task (function FINDCANDIDATE).

\textbf{Run-time complexity.} Algorithm~\ref{alg:ECDF} uses the schedulability tests in Proposition~\ref{prop:LC_schedulability} and Theorem~\ref{thm:improved_test}, both of which have pseudo-polynomial complexity. Function FINDCANDIDATE takes a constant amount of time for each $HC$ task in the list of potential candidates, and therefore its total complexity is linear in the size of the candidate list for each call. The list of potential candidates is initialized with all the $HC$ tasks, and in each iteration of the while loop, either a task is removed from this list or some $LC$ deadline is tightened by $1$ time unit. Further, when the $LC$ deadline cannot be reduced anymore, the task is also removed from the candidate list. Therefore, in the worst case, the $LC$ deadline of each $HC$ task is reduced until it reaches its $LC$ WCET, and in this case the while loop executes $\sum_{\tau_i \in \mathcal{H}_{\tau}} (D_i - C_i^L)$ number of times. Therefore, the overall complexity of Algorithm~\ref{alg:ECDF} is also pseudo-polynomial.

\subsection{Simulation results}
\label{sec:simulations}

\begin{figure}
\centering
%\subfigure[pCriticality = 0.3]{
%\includegraphics[width=0.3\linewidth]{figures/plot_deadline_normal_3.pdf}
%\label{fig:deadline_normal_0.3}
%}
\subfigure[pCriticality = 0.5]{
\includegraphics[width=0.72\linewidth]{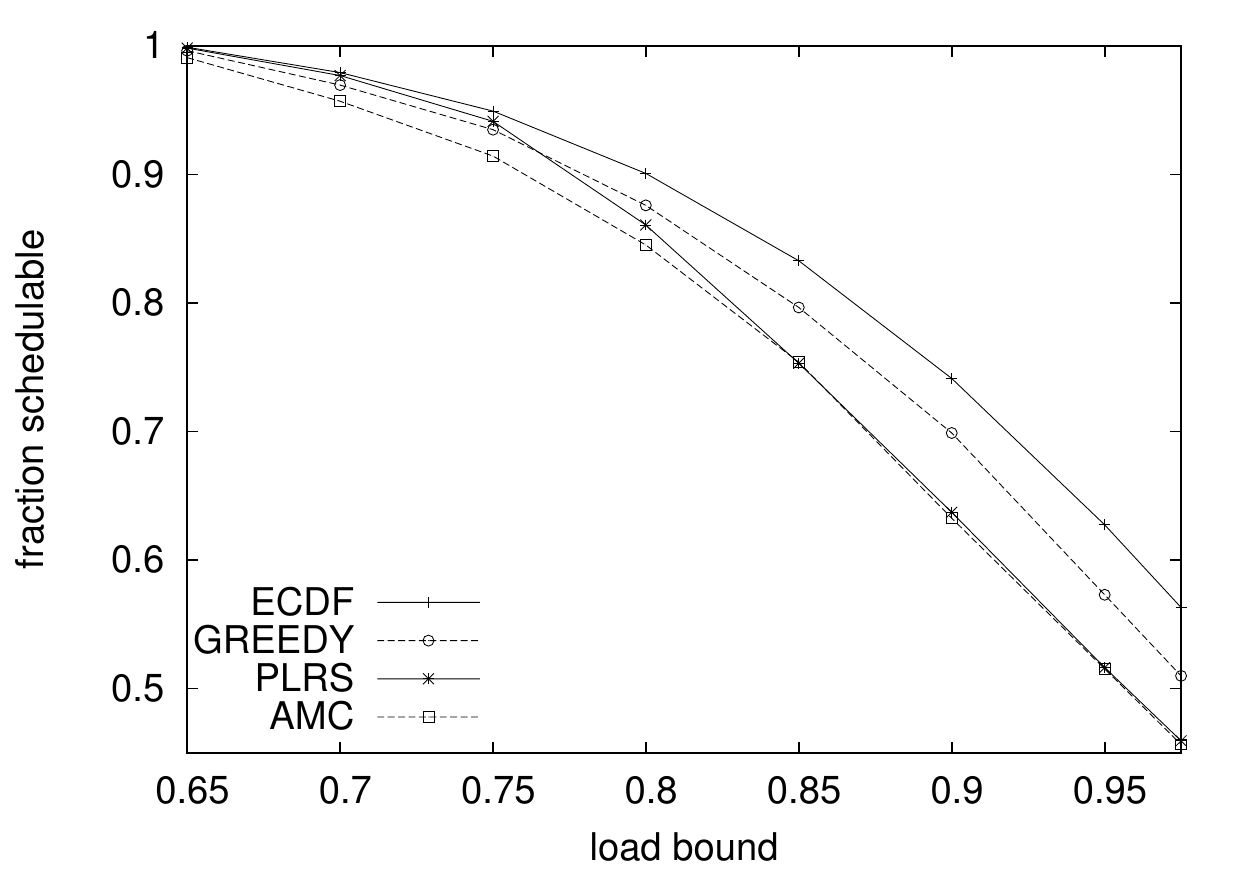}
\label{fig:deadline_normal_0.5}
}
\subfigure[pCriticality = 0.7]{
\includegraphics[width=0.72\linewidth]{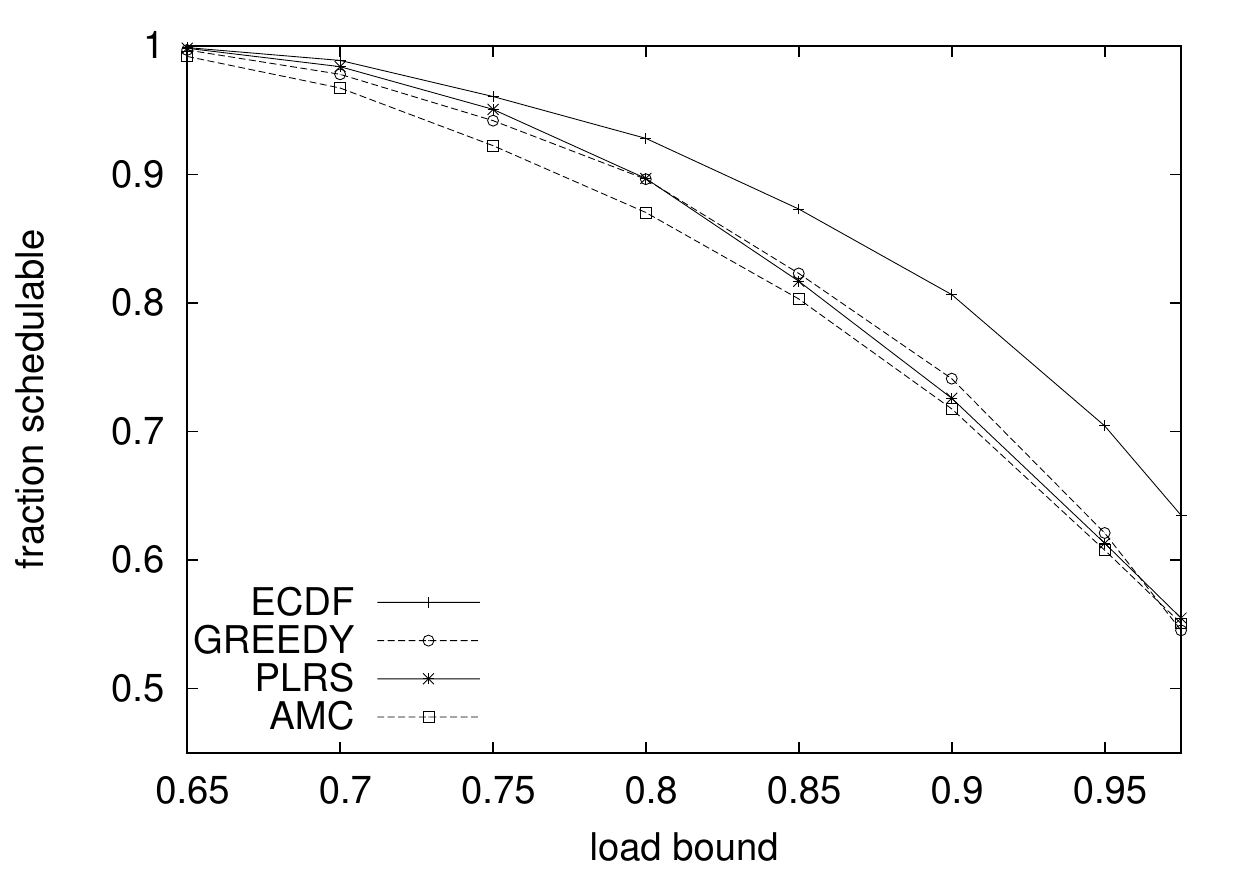}
\label{fig:deadline_normal_0.7}
}
\caption{Simulation results with deadlines in $[C_i^H, T_i]$}
\label{fig:deadline_normal}
\end{figure}

In this section we present simulation results comparing ECDF with AMC~\cite{BBD11}, PLRS~\cite{GES11}, and GREEDY~\cite{EkYi12}.

\begin{figure}
\centering
%\subfigure[pCriticality = 0.3]{
%\includegraphics[width=0.3\linewidth]{figures/plot_deadline_skewed_3.pdf}
%\label{fig:deadline_skewed_0.3}
%}
\subfigure[pCriticality = 0.5]{
\includegraphics[width=0.72\linewidth]{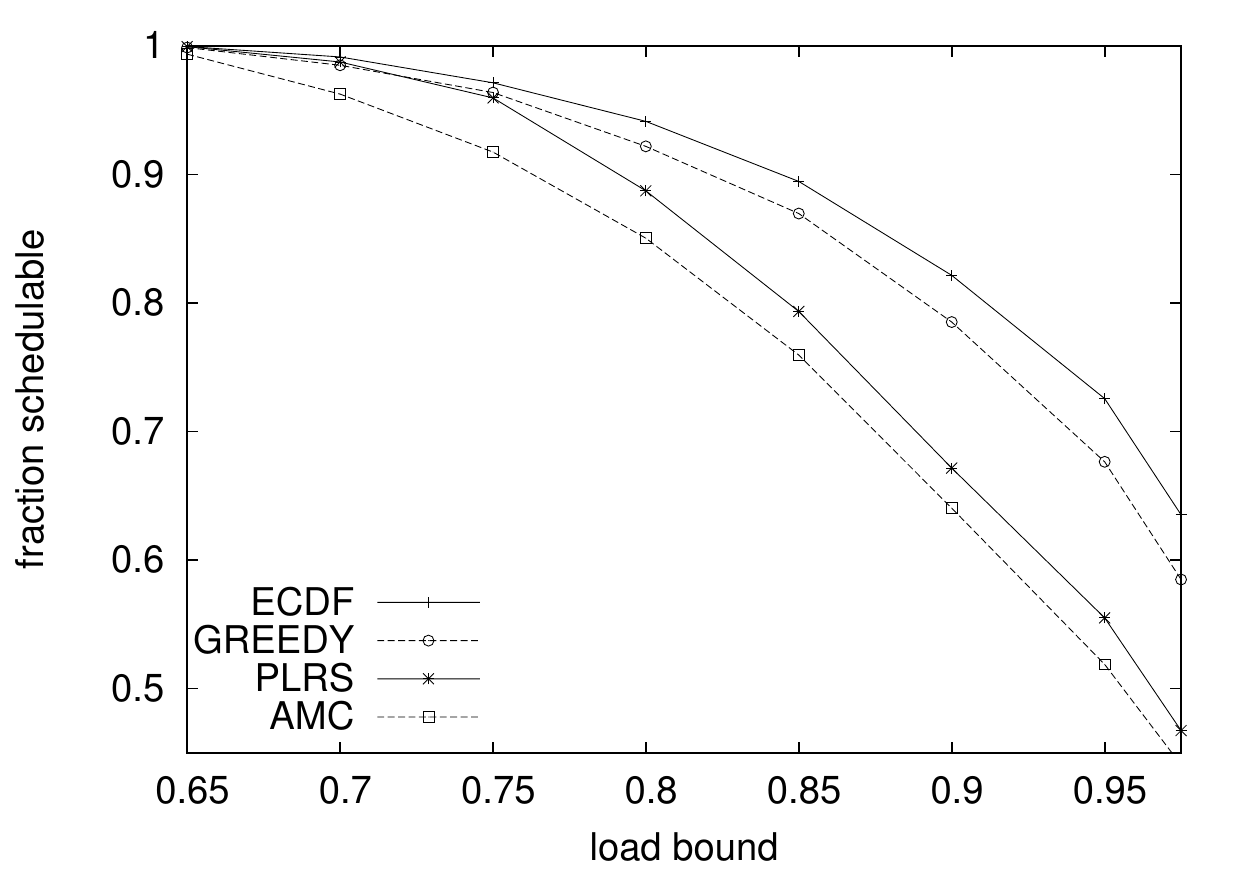}
\label{fig:deadline_skewed_0.5}
}
\subfigure[pCriticality = 0.7]{
\includegraphics[width=0.72\linewidth]{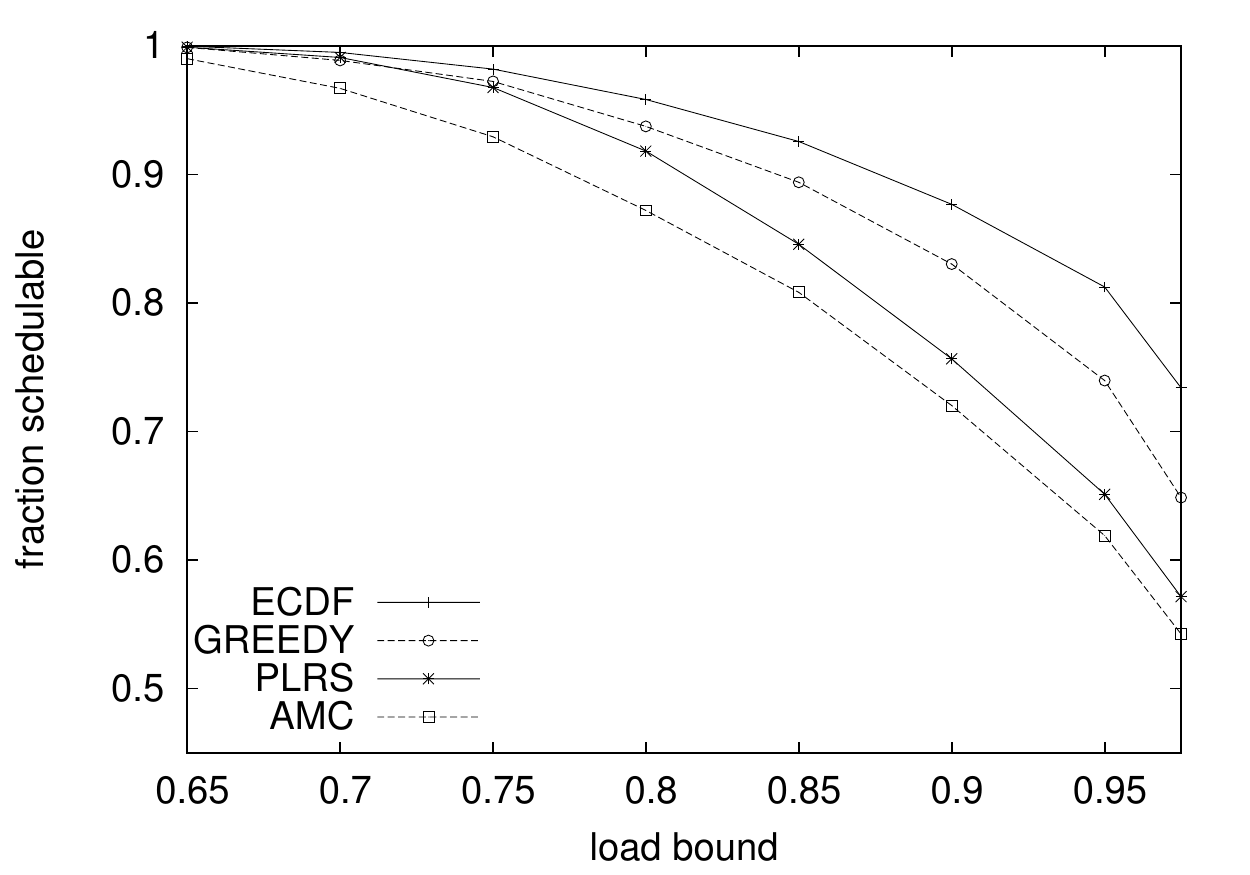}
\label{fig:deadline_skewed_0.7}
}
\caption{Simulation results with $HC$ deadlines in $[C_i^H + (T_i - C_i^H)/2, T_i]$}
\label{fig:deadline_skewed}
\end{figure}

\textbf{Simulation setting.} The various parameters are as follows.
\begin{itemize}
\item $T_i$ is drawn at random from $[5, 100]$.
\item $pCriticality$ denotes the probability that a task is a $HC$ task, and we choose values for this parameter from the set $\{ 0.5, 0.7 \}$.
\item $LC$ task utilization is drawn at random from $[0.02, 0.25]$. That is, if $T_i$ denotes the minimum separation, then its $LC$ WCET $C_i^L$ is drawn randomly from the range $[T_i*0.02, T_i*0.25]$.
\item Once the $LC$ WCET is fixed, if the task is a $HC$ task, then its $HC$ WCET $C_i^H$ is drawn at random from the range $[2*C_i^L,4*C_i^L]$.  
\item Task deadline $D_i$ is drawn at random either from $[C_i^H, T_i]$ (simulations in Figure~\ref{fig:deadline_normal}) or from $[C_i^H+(T_i-C_i^H)/2, T_i]$ (simulations in Figure~\ref{fig:deadline_skewed}).
\end{itemize} 
Tasks are generated using the above parameters one at a time until the following condition on system load is satisfied. 
{\scriptsize
\begin{equation*}
\max_{0 \leq t \leq t_{MAX}} \left \{ \frac{\max \left \{ \sum_{\tau_i \in \tau} dbf_i^L(t), \sum_{\tau_i \in \mathcal{H}_{\tau}} dbf_i^H(t) \right \}}{t} \right \} \leq lBound,
\end{equation*}
}
where $lBound \in \{ 0.65, 0.7, 0.75, 0.8, 0.85, 0.9, 0.95, 0.975 \}$, $dbf_i^L(t)$ is given by Equation~\eqref{eqn:dbfl}, and $dbf_i^H(t)$ is given by Equation~\eqref{eqn:dbfh}. This condition ensures that the load of the resulting task system in a purely $LC$ or $HC$ behavior does not exceed $lBound$. For each $pCriticality$ and $lBound$ values, we generated $10,000$ task sets and evaluated their schedulability using the four algorithms mentioned above. 

Figure~\ref{fig:deadline_normal} shows simulation results when task deadlines are drawn from the range $[C_i^H, T_i]$. In these figures, the x-axis denotes the value for $lBound$ and the y-axis plots the fraction of task sets deemed schedulable by the respective algorithms. The two figures are for different $pCriticality$ values, where a lower value denotes lower proportion of $HC$ tasks in the task system. As can be seen from these figures, ECDF clearly outperforms all the other algorithms in all scenarios, even in cases when the proportion of $HC$ tasks is small ($pCriticality = 0.5$). Further, this performance gap widens with increasing system load and proportion of $HC$ tasks. When the proportion of $HC$ tasks is higher, there is more opportunity for ECDF to tighten task deadlines, and therefore we can see the improved performance. An interesting observation is that GREEDY which is also based on a deadline tightening strategy, does not show a similar improvement with increasing proportion of $HC$ tasks. We suspect this is mainly because of the large pessimism in the schedulability test used by the algorithm, and the benefits of reducing this pessimism is clearly seen in the case of ECDF.

Figure~\ref{fig:deadline_skewed} shows simulation results when $HC$ task deadlines are drawn from the range $[C_i^H + (T_i - C_i^H)/2, T_i]$ and $LC$ task deadlines are drawn from the range $[C_i^H, T_i]$. Since $HC$ task deadlines are larger when compared to the earlier simulations, deadline tightening strategies ECDF and GREEDY have more opportunities to tighten $HC$ task deadlines in these systems. As can be seen from the figures, these algorithms indeed significantly outperform AMC and PLRS in this scenario. An even more interesting observation is that ECDF continues to outperform GREEDY, and this performance gap widens with increasing system load and proportion of $HC$ tasks.

\begin{figure}
\centering
%\subfigure[pCriticality = 0.3]{
%\includegraphics[width=0.3\linewidth]{figures/plot_deadline_skewed_3.pdf}
%\label{fig:deadline_skewed_0.3}
%}
\subfigure[$4$ tasks per set]{
\includegraphics[width=0.72\linewidth]{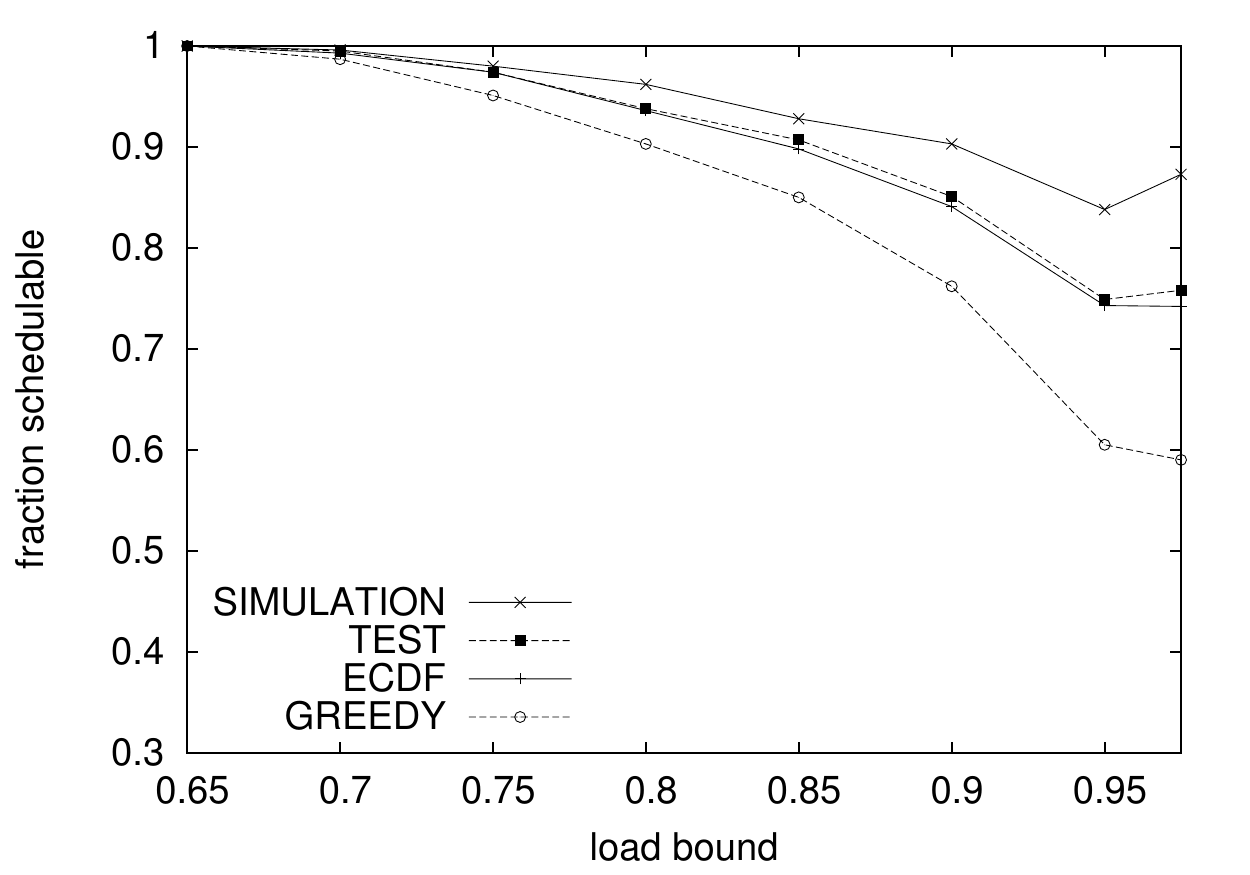}
\label{fig:deadline_skewed_exhaustive_0.74}
}
\subfigure[$6$ tasks per set]{
\includegraphics[width=0.72\linewidth]{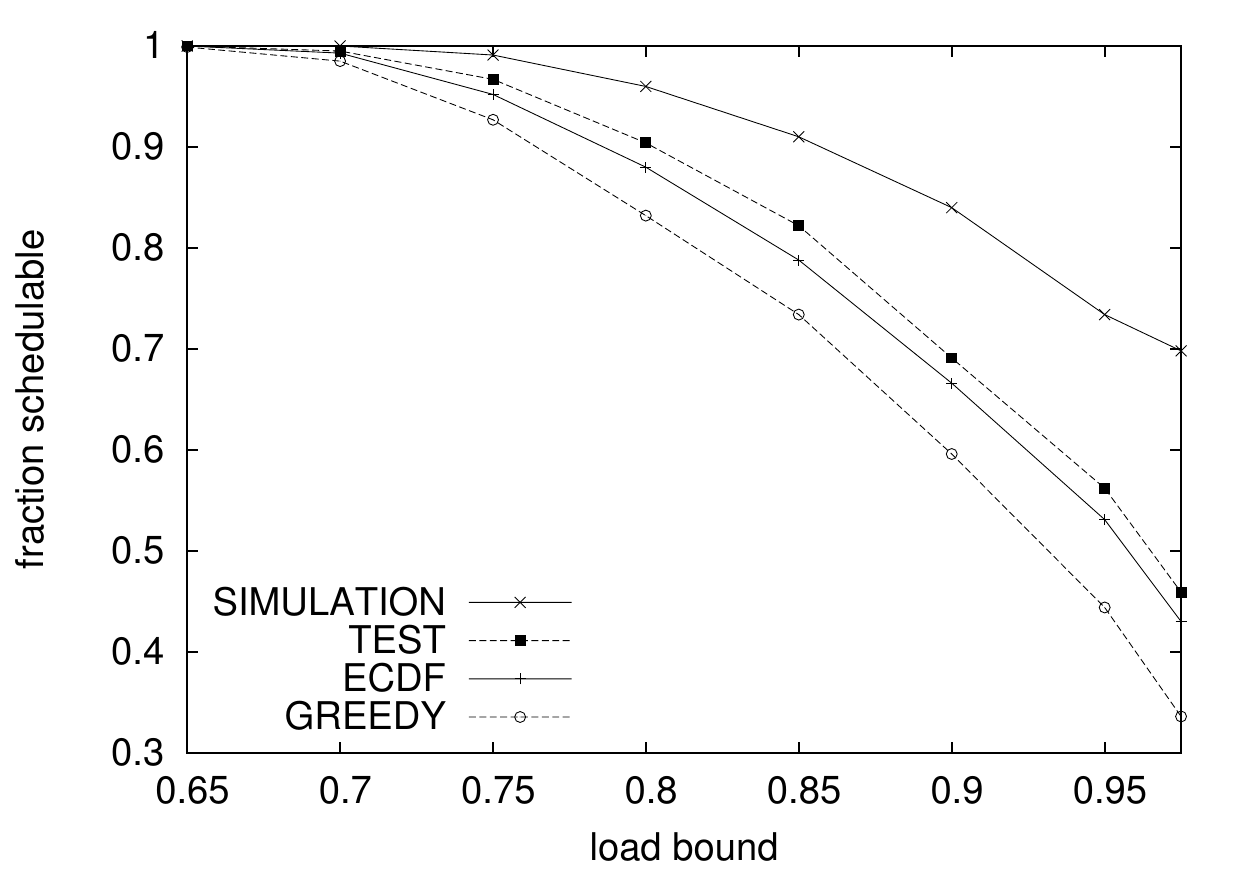}
\label{fig:deadline_skewed_exhaustive_0.76}
}
\caption{Simulation results for comparison with exhaustive search}
\label{fig:deadline_skewed_exhaustive}
\end{figure}

We also did experiments to evaluate how well ECDF performs in comparison to exhaustive deadline search (see Figure~\ref{fig:deadline_skewed_exhaustive}). We considered two exhaustive strategies for this comparison. The first one, denoted SIMULATION, tries all possible tightened deadline values for the $HC$ tasks and then evaluates schedulability by simulating EDF strategy for all possible values of $t_1$. The second one, denoted TEST, also tries all possible tightened deadline values for the $HC$ tasks, but evaluates schedulability using Theorem~\ref{thm:improved_test}. To make these exhaustive simulations practically feasible, we only considered task sets with a small number of tasks ($4$ or $6$), limited the range of $T_i$ to $[10, 30]$, and generated $1000$ task sets for each $pCriticality$ and $lBound$ values.

Figure~\ref{fig:deadline_skewed_exhaustive} shows the simulation results for GREEDY, ECDF, TEST, and SIMULATION, when $pCriticality = 0.7$ and $HC$ task deadlines are chosen from $[C_i^H+(T_i-C_i^H)/2, T_i]$. Figure~\ref{fig:deadline_skewed_exhaustive_0.74} (likewise Figure~\ref{fig:deadline_skewed_exhaustive_0.76}) shows the results when number of tasks in the task set is $4$ (likewise $6$). As can be seen in both the cases, ECDF performs almost as well as TEST, suggesting that the deadline search heuristic is very effective. The gap between SIMULATION and TEST indicates the pessimism still present in the improved test, and as expected this gap widens with increasing number of tasks.     

%\textbf{Discussion.} Although we performed simulations only on constrained deadline task systems, we expect this performance gap between ECDF and the other algorithms to further widen when tasks have deadline greater than minimum separation. This is reasonable given similar trends when task deadlines were chosen from the range $[C_i^H + (T_i - C_i^H)/2, T_i]$ as opposed to from the range  $[C_i^H, T_i]$ (Figure~\ref{fig:deadline_skewed} versus Figure~\ref{fig:deadline_normal}). 
